%% file: paper.tex
\documentclass{fairmeta}
\usepackage{amsmath,amssymb,booktabs,array,tabularx,graphicx,microtype,enumitem,hyperref,url}
\DeclareMathOperator*{\argmax}{arg\,max}
\usepackage{wrapfig,tcolorbox,capt-of,placeins,needspace}
\usepackage{pgfplots}
\pgfplotsset{compat=1.18}

\usepackage{hyperref}
\usepackage{url}

\usepackage{graphicx}
\usepackage{wrapfig}
\usepackage{subcaption}
\usepackage{booktabs}
\usepackage{multirow}
\usepackage{lmodern}

\usepackage{longtable}

\usepackage[table]{xcolor}   
\usepackage{pifont}
\usepackage{xcolor}
\usepackage{xcolor}
\definecolor{metablue}{HTML}{0081FB}
\colorlet{metalight}{metablue!15}   
\usepackage{bbm}
\newcommand{\ind}{\mathbbm{1}}

\usepackage{caption}
\crefname{figure}{Fig.}{Figs.}
\Crefname{figure}{Fig.}{Figs.}
\crefname{table}{Tab.}{Tabs.}
\Crefname{table}{Tab.}{Tabs.}
\crefname{section}{Sec.}{Secs.}
\Crefname{section}{Sec.}{Secs.}
\crefname{equation}{Eq.}{Eqs.}
\Crefname{equation}{Eq.}{Eqs.}

\newcommand{\cmark}{\textcolor{green!60!black}{\ding{51}}}
\newcommand{\xmark}{\textcolor{red}{\ding{55}}}

\title{ATTEST: Evidence-Grounded Reinforcement Learning for Long-Audio Temporal Understanding}
\author[1,2,*]{Ashutosh Chaubey}
\author[1]{Jinxi Guo}
\author[1]{Minh Tran}
\author[1]{Michael Andersland}
\author[1]{Ozlem Kalinli}
\affiliation[1]{Meta Superintelligence Labs}
\affiliation[2]{University of Southern California}
\contribution[*]{Work done during an internship at Meta}
\input{secs/0_abstract}
\date{October 12, 2026}
\begin{document}
\raggedbottom
\maketitle

\begin{figure}[t]
  \centering
  \begin{subfigure}[b]{0.58\textwidth}
    \centering
    \includegraphics[width=\linewidth]{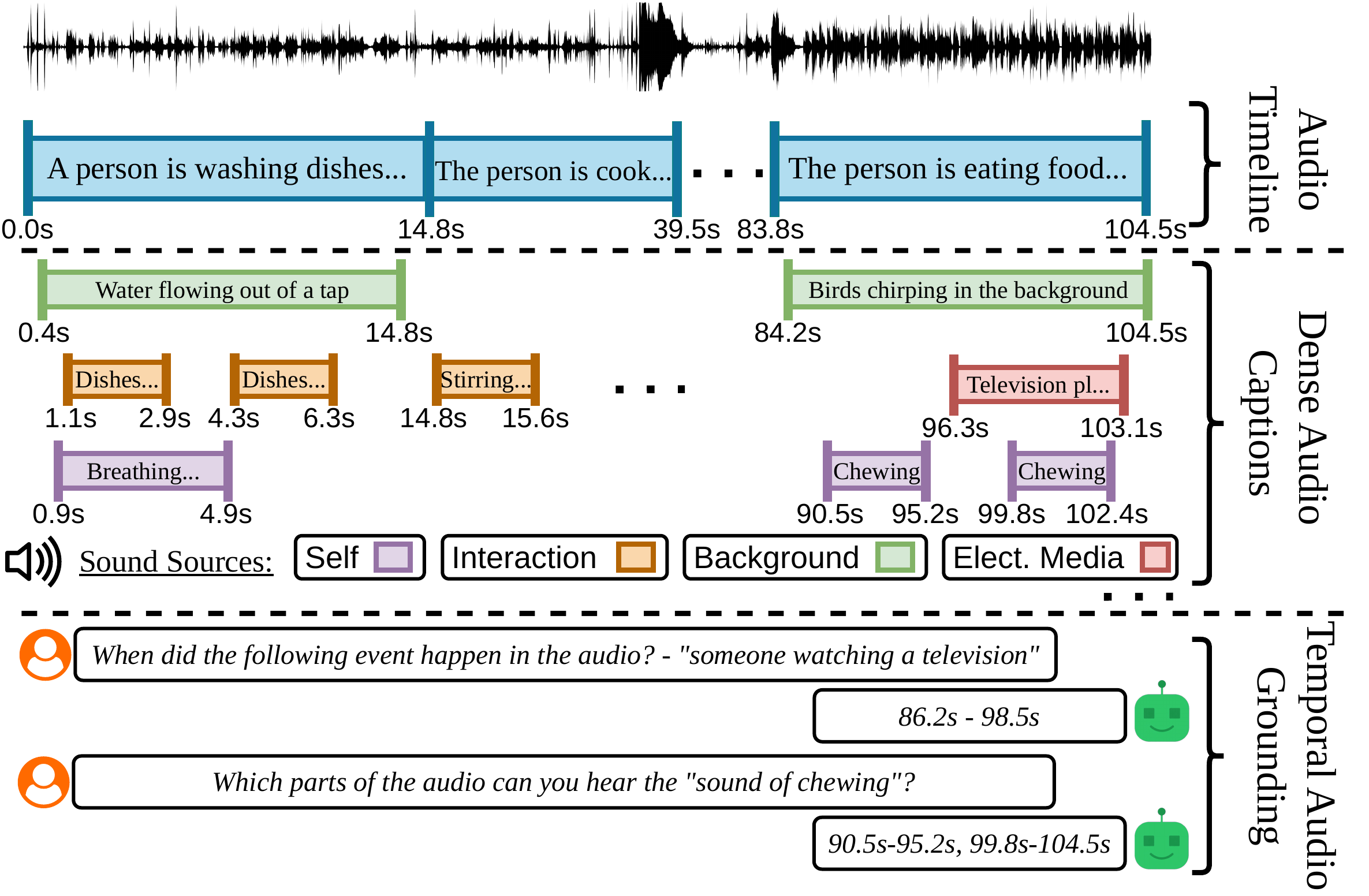}
    \label{fig:task_samples_tasks}
  \end{subfigure}\hfill
  \begin{subfigure}[b]{0.40\textwidth}
    \centering
    \includegraphics[width=\linewidth]{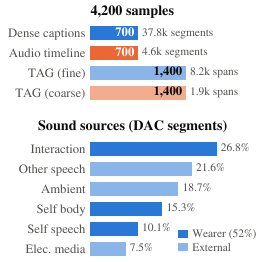}
    \label{fig:task_samples_stats}
  \end{subfigure}
  \caption{\textbf{EgoSCAPE benchmark overview.} \textbf{(Left)} The three benchmark tasks: audio timeline generation, dense audio captioning (DAC) with sound sources, and temporal audio grounding (TAG). \textbf{(Right)} Samples per task and sound-source distribution.}
  \label{fig:task_samples}
\end{figure}

\input{secs/1_intro}
\input{secs/2_related}
\input{secs/3_data}
\input{secs/4_method}

\input{secs/5_experiments}

\input{secs/6_conclusion}

\clearpage

\bibliography{paper}
\bibliographystyle{assets/plainnat}

\input{secs/A_appendix}

\end{document}

%% file: secs/0_abstract.tex
\abstract{
Temporal grounding is fundamental to audio understanding, yet large audio language models (LALMs) remain weak at it on long audio, where hallucinations compound and annotated data is scarce. Existing benchmarks cover only short clips or offer coarse, synthetic annotations. We introduce \textbf{EgoSCAPE}, a manually verified benchmark created from Ego4D with 700 real audios of durations up to 5 minutes, annotated with dense captions, sound sources, audio timelines, and temporal grounding queries. We further propose \textbf{ATTEST}, a GRPO-based reinforcement learning framework that improves temporal grounding without changing the base model's architecture or token vocabulary. ATTEST combines (i) a counterfactual grounding reward penalizing predictions that survive removal of their acoustic evidence, (ii) an attention-density reward encouraging generation to attend to its own claimed time spans, and (iii) token-wise advantages giving segment-level credit for long dense captions. ATTEST outperforms prior temporal grounding methods on existing dense captioning and grounding benchmarks, with fewer hallucinations. Together, EgoSCAPE and ATTEST establish a rigorous evaluation–training recipe that brings temporally grounded audio understanding closer to practical, long-form deployment.
}

%% file: secs/1_intro.tex
\section{Introduction}

Temporal grounding is fundamental for building audio models for complex audio understanding and causal reasoning. \emph{Sound event detection} (SED) deals with detecting various sound events densely within an audio along with their onset and offset timestamps \citep{mesaros2016metrics_sed,mesaros2017dcase,mesaros2025decade_sed}. \emph{Language-based audio moment retrieval} (also known as \emph{temporal audio grounding} or TAG) deals with identifying where an audio query described by text occurs in an audio clip \citep{martin2023strong,munakata2025language_clotho_moment_lamr,munakata2026castella}. With the advent of large audio language models (LALMs) \citep{chu2024qwen2audio,tang2024salmonn,ghosh2025audioflamingo2_af2,ghosh2026audioflamingonext_afnext,ding2025kimiaudio}, more recently, there has been an increased interest in \emph{dense audio captioning} (DAC) which involves captioning all the audio events occurring in an audio along with their timestamps \citep{kumar2026tac,shao2026listening_lat_audio,wang2026listening_timeaudio_ftar}. Such dense captioning requires both semantic and temporal understanding about the input audio necessary for downstream reasoning in LALMs. 

Prior dense audio captioning work \citep{wang2026listening_timeaudio_ftar,shi2026towards_timepro_rl} targets short 10–30 second clips from datasets such as AudioSet \citep{jort2016audioset,shawn2021audioset_sl}, AudioTime \citep{xie2025audiotime}, and TACOS \citep{primus2025tacos}. Some studies generate temporal annotations for longer audio synthetically, but such evaluations may not reflect real-world use \citep{munakata2025language_clotho_moment_lamr,kumar2026tac,sun2026spotsound}. LAT-Bench \citep{shao2026listening_lat_audio} covers audio up to 30 minutes but provides coarse event timelines rather than dense captions (see \cref{fig:task_samples}), while CASTELLA \citep{munakata2026castella} spans up to 5 minutes with only coarse temporal grounding annotations.



On the modeling side, existing approaches to temporal understanding with LALMs either modify the base model's input token representation \citep{shi2026towards_timepro_rl,kutsakov2026gigachat} or alter its architecture and output vocabulary \citep{wang2026listening_timeaudio_ftar}. These changes consume extra tokens, increasing inference cost and context usage, and frequently cause catastrophic forgetting on tasks where the base model originally excelled \citep{sun2026spotsound,kumar2026tac,shi2026towards_timepro_rl,kutsakov2026gigachat}. Furthermore, large language models (LLMs) and LALMs are prone to hallucination \citep{huang2025survey_llm_hallucination,kuan2024understanding_audio_hallucination,cheng2026ahabench,chaubey2026avere,Chaubey_2026_CVPR_moddpo,pang2026do_voxparadox}, which worsens in long-context settings \citep{hsieh2024ruler}, a problem further compounded for long-audio temporal grounding by the scarcity of data noted in the above paragraph.

To address these gaps, we introduce \textbf{EgoSCAPE}, a long-audio temporal grounding benchmark built on the Ego4D dataset \citep{grauman2022ego4d}, with manually verified dense audio captions, coarse audio timelines, and temporal grounding annotations (see \cref{fig:task_samples}). It comprises 700 audio clips of up to 5 minutes each, totaling 13.15 hours. Beyond temporal captions of acoustic events, we also annotate each sound's source into one of the six categories mentioned in \cref{fig:task_samples}.

To improve long-audio temporal understanding in LALMs, we propose \textbf{ATTEST}, a reinforcement learning framework that mitigates hallucinations and promotes encourages predictions grounded in the audio without altering the base model's architecture or token vocabulary. ATTEST introduces three components: (i) a counterfactual reward that ensures predictions to be grounded in the acoustic events of the corresponding temporal region, (ii) an attention density reward that steers the model's attention toward the audio tokens of the time span each prediction claims during generation, and (iii) a token-wise advantage that provides fine-grained supervision for long, dense captions. Experiments on existing dense captioning and temporal grounding benchmarks, as well as EgoSCAPE, show that ATTEST outperforms prior temporal grounding methods by significant margins while producing hallucination-free predictions.

In summary, our main contributions are: (i) \textbf{EgoSCAPE}, a new benchmark for dense audio captioning, audio timeline generation, and temporal audio grounding, built from a subset of Ego4D \citep{grauman2022ego4d} with audio of up to 5 minutes; (ii) \textbf{ATTEST}, a novel RL framework that mitigates hallucinations and aligns the model's attention with the time spans it predicts; and (iii) extensive experiments showing that ATTEST outperforms existing baselines on dense audio captioning and temporal audio grounding without regressing the base model's performance on general audio understanding.

%% file: secs/2_related.tex
\section{Related Works}


\noindent \textbf{Large audio language models and long audio.} LALMs augment text-only LLMs by accepting an audio input through an audio encoder \citep{radford2023robust_whisper,chen2023beats} and training on a large audio-text corpus \citep{tang2024salmonn,chu2024qwen2audio,kong2024audioflamingo_af1,ghosh2025audioflamingo2_af2,ghosh2026audioflamingo3_af3,ghosh2026audioflamingonext_afnext}. Moreover, omni LLMs extend this recipe to accept multiple modalities including audio and text, showing better performance than audio-only models on various benchmarks \citep{xu2025qwen25omnitechnicalreport,xu2025qwen3omni}. Evaluation of these models has largely been using short audio benchmarks such as MMAR \citep{ma2026mmar}, MMAU \citep{sakshi2025mmau} and MMSU \citep{wang2026mmsu}. There has been recent focus on long-audio processing where benchmarks such as AudioMarathon \citep{he2025audiomarathon}, BLAB \citep{ahia2025blab}, and VoiceGiraffe \citep{ye2026voicegiraffe} show challenges to long audio processing and AudioFlamingo series of models \citep{ghosh2025audioflamingo2_af2,ghosh2026audioflamingo3_af3,ghosh2026audioflamingonext_afnext} support long audios by training on large-scale synthetic datasets. Our benchmark and proposed RL framework tackle the specific challenge of long-audio temporal grounding and understanding. 

\begin{figure}
    \centering
    \includegraphics[width=\linewidth]{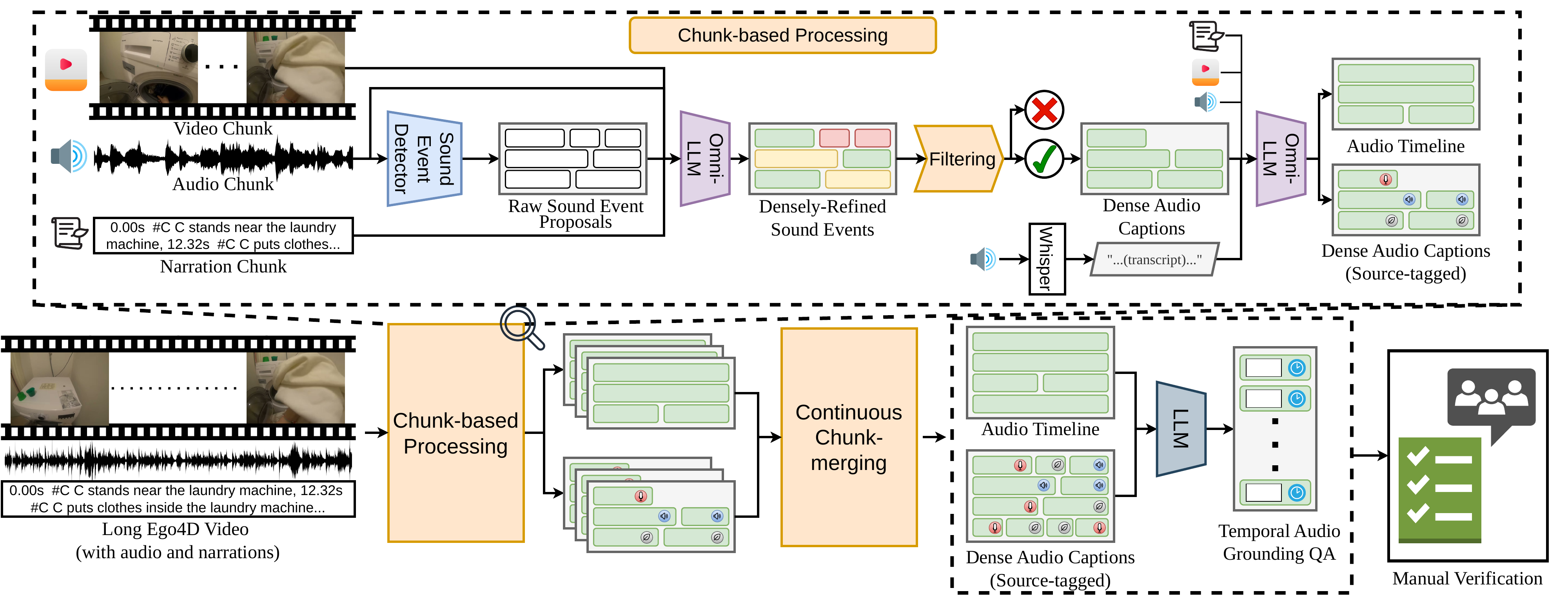}
    
    \vspace{-1em}
    
    \caption{\textbf{EgoSCAPE benchmark creation pipeline.} To ensure high-quality annotations, we first chunk the audios that are automatically annotated using an omni-LLM with additional supervision from the video and narrations available for Ego4D. All the generated samples for the benchmark are manually verified and corrected at the end.}
    \label{fig:egoscape_benchmark_pipeline}
    \vspace{-1em}
\end{figure}


\noindent \textbf{Temporal audio grounding.} While temporal grounding tasks such as sound event detection have long been prominent \citep{mesaros2017dcase,mesaros2025decade_sed,munakata2025language_clotho_moment_lamr,munakata2026castella}, use of multimodal LLMs for temporal grounding and dense captioning is fairly recent \citep{ghosh2024compar,wang2026listening_timeaudio_ftar,shao2026listening_lat_audio,shi2026towards_timepro_rl}. TimeAudio \citep{wang2026listening_timeaudio_ftar} introduces temporal markers in the model vocabulary to improve DAC and TAG for short audios. TimePro-RL \citep{shi2026towards_timepro_rl} uses audio-side time tokens and RL to improve DAC and TAG for short utterances. SpotSound \citep{sun2026spotsound} and TAC \citep{kumar2026tac} utilize SFT training over synthetic datasets to improve performance on TAG and DAC respectively. While most of these baselines deal with short audios of around 30 seconds, LAT-Audio \citep{shao2026listening_lat_audio} introduces an agentic framework for audio timeline generation and TAG for audios up to 30 minutes. GigaChat-Audio \citep{kutsakov2026gigachat} performs TAG on speech up to 2 hours. Compared to these approaches, EgoSCAPE benchmark consists of DAC with sound source, Audio Timeline, and TAG annotations for audios up to 5 minutes. Moreover, the proposed ATTEST framework improves temporal understanding without modifying model vocabulary or consuming additional input context via time-tokens.

%% file: secs/3_data.tex
\section{EgoSCAPE - Egocentric Sound CAPtion of Events}
\label{sec:egoscape_data}


We describe the benchmark creation pipeline for EgoSCAPE in this section as shown in \cref{fig:egoscape_benchmark_pipeline}. Compared with other benchmarks of temporal audio grounding (see \S\ref{appsubsec:benchmark_comparison}), our benchmark is rich in annotations as it contains dense annotation over long audio clips.


\subsection{Automatic Benchmark Creation}
\label{subsec:automatic_benchmark_creation}
Traditional SED models \citep{chen2023beats,shao2024pretrained_atst_sed} predict accurate temporal boundaries but are restricted to a fixed label set (typically AudioSet \citep{jort2016audioset} or ESC-50 \citep{piczak2015dataset_esc50}), limiting annotation richness. Omni-LLMs, in contrast, generate open-ended annotations but their temporal boundaries often drift over long multimodal inputs \citep{shao2026listening_lat_audio}. To combine their strengths, we split Ego4D videos into 30-second chunks, obtain SED labels over AudioSet classes using a BEATs-based detector \citep{chen2023beats}, and use an omni-LLM to verify and refine them (prompt in \cref{tab:prompt_sed_verify}), discarding chunks where over 30\% of events are marked incorrect. A second omni-LLM pass annotates sound sources and generates a coarse audio timeline (prompts in \cref{tab:prompt_densesource,tab:prompt_audio_timeline_generation}), incorporating Whisper transcripts \citep{radford2023robust_whisper} for speech content. Both passes also take the Ego4D video and textual narrations as input to help disambiguate events and sources.

We then merge consecutive chunk annotations into a single annotation per audio file, using an LLM to correctly join events spanning chunk boundaries in the timelines (prompt in \cref{tab:prompt_timeline_rewrite}). Finally, an LLM generates temporal grounding queries: fine-grained queries about atomic events from the dense captions, and coarse queries about multi-sound acoustic scenes from the timelines (prompts in \cref{tab:prompt_tag_fine_grained,tab:prompt_tag_coarse}). We use Qwen3-Omni-30B-A3B-Instruct \citep{xu2025qwen3omni} as the omni-LLM and Qwen3-235B-A22B-Instruct \citep{yang2025qwen3} as the LLM throughout the pipeline. 

\subsection{Human Verification and Benchmark Stats}

To ensure high data quality, each generated annotation in EgoSCAPE is verified by three human annotators (details in \S\ref{appsubsec:human_verification_details}), and annotations flagged as incorrect are corrected accordingly. The final benchmark comprises 700 unique audio clips drawn from 121 unique Ego4D videos, totaling approximately 13.15 hours. Further benchmark statistics are provided in \S\ref{appsubec:benchmark_stats}.

%% file: secs/4_method.tex
\section{ATTEST: ATTention and Evidence-grounded Segment Training}
\label{sec:attest_method}

We propose \textbf{ATTEST}, an RL-based framework that modifies Group Relative Policy Optimization (GRPO) \citep{shao2024deepseekmath_grpo} to improve temporal audio grounding over long audios as described in this section.
We perform RL only on the dense audio captioning and audio timeline generation tasks and show that the temporal grounding capabilities transfer to other tasks as well in \S\ref{subsec:results_grounding} \& \S\ref{subsec:results_general_benchmarks}.


\begin{wrapfigure}{r}{0.45\textwidth}
    \vspace{-2em}
    \centering
    \includegraphics[width=0.43\textwidth]{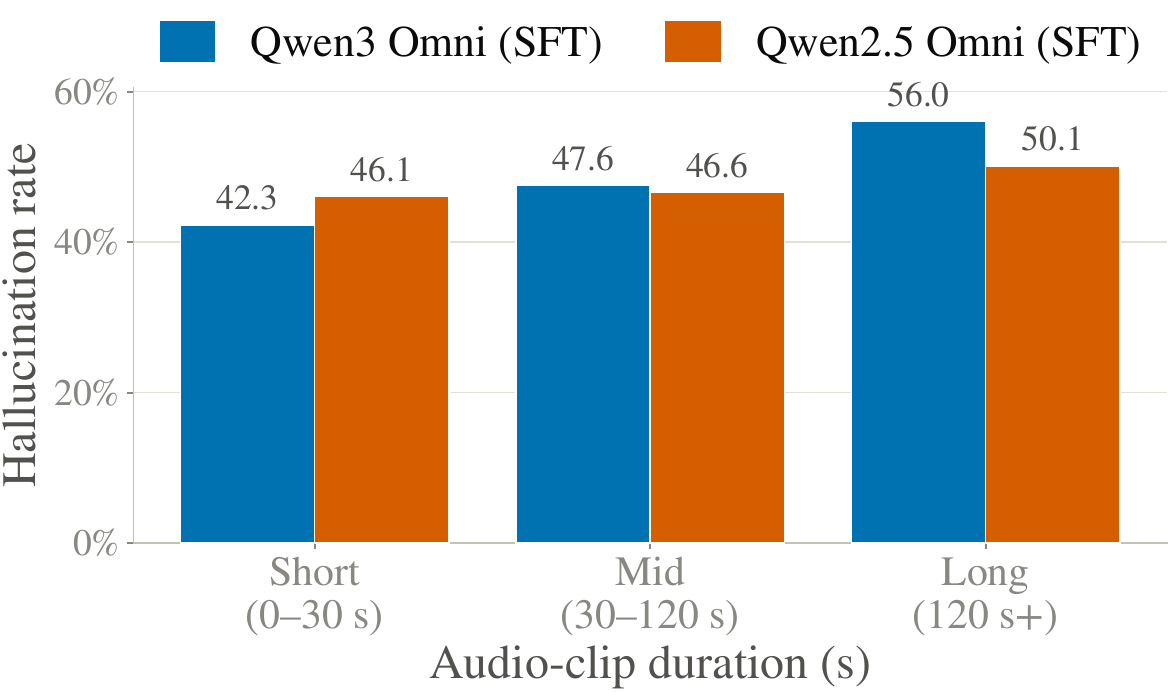}
    \caption{Hallucination for difft. audio durations in EgoSCAPE (details - \S\ref{appsubsec:figure3_details}).}
    \label{fig:hallucination_sft}
    \vspace{-2em}
\end{wrapfigure}
\textbf{Notation.}
Let $a$ denote the input audio of duration $T_{\text{aud}}$ and $x$ the full input token sequence (audio tokens $\mathcal{T}_{\text{aud}}$ and text prompt tokens $\mathcal{T}_{\text{tex}}$). The ground truth is a set of $N_g$ segments (or triplets) $D=\{(s_i,e_i,c_i)\}_{i=1}^{N_g}$, where $(s_i,e_i)$ are the start/end timestamps of an event described by caption $c_i$. For a rollout $g$ sampled from $\pi_{\theta_{\text{old}}}(\cdot\mid x)$, the prediction is $\hat{D}^{(g)}=\{(\hat{s}^{(g)}_j,\hat{e}^{(g)}_j,\hat{c}^{(g)}_j)\}_{j=1}^{N^{(g)}_p}$, decoded from the fixed output format \emph{``0.1 - 2.2, person speaking, 1.4 - 3.8, birds chirping, \dots''}. We write $\mathcal{G}^{(g)}_j$ for the set of generated token positions that spell out the $j$-th predicted triplet, and $\mathcal{T}(s,e)\subseteq\mathcal{T}_{\text{aud}}$ for the audio tokens whose receptive field overlaps the time interval $[s,e]$. $\mathrm{Sim}(\cdot,\cdot)\in[0,1]$ is the FENSE caption score \citep{zhou2022can_fense} and $\mathrm{tIoU}$ the temporal intersection-over-union. Throughout, a superscript $(g)$ marks a quantity belonging to rollout $g$; we suppress it inside expressions that are defined for a single generic rollout.

\subsection{Counterfactual Audio-Grounding (CAG)}
\label{subsec:cag}

Our preliminary experiments on EgoSCAPE (\cref{fig:hallucination_sft}) show that the hallucination rate (defined at \S\ref{appsubsubsec:captioning_metrics}) rises on longer audio clips. This suggests that the model is increasingly recovering plausible events from linguistic and dataset priors instead of grounding events in the audio. To mitigate this behavior, we propose the following intervention and reward.

\textbf{Intervention.} For each training sample we modify a target fraction $p$ of the clip. We draw $K$ non-overlapping windows $\mathcal{W}=\{w_1,\dots,w_K\}$, $w=(t^{w}_{ms},t^{w}_{me})$, uniformly at random in time.
We apply a corruption operator $M(\cdot)$ to obtain $a'$ such that the original audio events in $\mathcal{W}$ are no longer present. We then decode a \emph{single} greedy counterfactual reference $\hat{D}'=\pi_{\theta_{\text{old}}}(a')$, which is shared across the GRPO rollout group, and score each of the $G$ sampled rollouts $\hat{D}^{(g)}$ on the original audio $a$ against it as described below.

\textbf{Reward.} Let $\hat{D}^{(g)}_w=\{\hat{d}^{(g)}_j : [\hat{s}^{(g)}_j,\hat{e}^{(g)}_j] \subseteq w\}$ where $\hat{d}^{(g)}_j$ is the segment triplet $(\hat{s}^{(g)}_j,\hat{e}^{(g)}_j,\hat{c}^{(g)}_j)$ and $\subseteq$ denotes containment of the predicted support in the window. 
Similarly, we can define $\hat{D}'_w$ for the counterfactual reference. For each $j\in\hat{D}^{(g)}_w$ we measure how strongly the prediction \emph{survives} the intervention and penalize it as follows:
\begin{equation}
\label{eq:cag}
\small
\chi^{(g)}_j = \underbrace{ - \max_{\hat{d}' \in \hat{D}'_w} \Big[ \text{tIoU}\big(\hat{d}^{(g)}_j, \hat{d}'\big) \cdot \text{Sim}\big(\hat{d}^{(g)}_j, \hat{d}'\big) \Big]}_{\text{segment-level reward}}
\end{equation}
Averaging the per-segment penalties within a window gives the window score $q^{(g)}_w$, and averaging over the $K$ windows gives the response-level reward:
\begin{equation}
\label{eq:cag_agg}
q^{(g)}_w \;=\; \frac{1}{\max\big(1,|\hat{D}^{(g)}_w|\big)}\sum_{j\in \hat{D}^{(g)}_w} \chi^{(g)}_j,
\qquad
R^{(g)}_{\text{CAG}} \;=\; \frac{1}{K}\sum_{w\in\mathcal{W}} q^{(g)}_w .
\end{equation}
$\chi^{(g)}_j\in[-1,0]$ varies across rollouts even though $\hat{D}'$ is fixed, since each rollout agrees with the counterfactual reference to a different degree within each window. We found the silence operator to work best for $M(\cdot)$ as mentioned in \S\ref{subsubsec:hyperparam_reward_design}.

\subsection{Attention Density Grounding}
\label{subsec:ad}

Because generation is autoregressive and the output format is fixed, the model should attend more strongly to the audio tokens $\mathcal{T}(\hat{s}_j, \hat{e}_j)$ while it emits the triplet for $[\hat{s}_j, \hat{e}_j]$. We verify this empirically in \S\ref{subsec:attention_behavior}. Among samples matched in segment count and duration, those with higher FENSE show consistently higher attention density on their claimed spans.


Let $\bar{A}_{t,u}$ be the attention weight from generated token $t$ to input token $u$. We define the attention density of predicted segment $j$ as the fraction of audio-directed attention mass that falls inside its own temporal support, normalized by the fraction $f_j=(\hat{e}_j-\hat{s}_j)/T_{\text{aud}}$ of the audio clip that the segment occupies, and the corresponding response-level reward as the mean density over the predicted segments:
\begin{equation}
\label{eq:ad}
\small
\rho_j \;=\; \underbrace{ \frac{1}{f_j}\cdot\frac{1}{|\mathcal{G}_j|}\sum_{t\in\mathcal{G}_j}
\frac{\sum_{u\in\mathcal{T}(\hat{s}_j,\hat{e}_j)} \bar{A}_{t,u}}
     {\sum_{u\in\mathcal{T}_{\text{aud}}} \bar{A}_{t,u}},}_{\text{segment-level reward}}
\qquad
R_{\text{AD}} \;=\; \frac{1}{N_p}\sum_{j=1}^{N_p} \rho_j .
\end{equation}
 
To ensure that the model does not hack  $R_\text{AD}$ by predicting short segments, we cap $\rho$ to a maximum of 5. We use the last layer's attention averaged across all heads to compute $\bar{A}_{t,u}$, based on empirical findings in \cref{tab:r_ad_attention_layer} (\S\ref{subsubsec:hyperparam_reward_design}). 

\paragraph{Correctness Reward ($R_{\text{DAC}}$).} In addition to the above rewards, we also apply a correctness reward for dense captioning which directly utilizes the ground-truth annotations. For each rollout $g$ we solve a Hungarian assignment between
$\hat{D}^{(g)}$ and $D$ maximizing
\begin{equation}
\label{eq:hungarian_matching_score}
\mathrm{Score}(i,j) \;=\; \alpha\,\mathrm{tIoU}\big((s_i,e_i),(\hat{s}_j,\hat{e}_j)\big)
\;+\;(1-\alpha)\,\mathrm{Sim}(c_i,\hat{c}_j),
\end{equation}
yielding an assignment $\sigma^{(g)}(j)\in\{0,1,\dots,N_g\}$, where $\sigma^{(g)}(j)=0$ marks an
unmatched (spurious) prediction. The segment-level reward is the matched payoff itself,
\begin{equation}
\label{eq:dac_seg}
\small
r^{(g)}_j \;=\;
\begin{cases}
\alpha\,\mathrm{tIoU}\big((s_{\sigma^{(g)}(j)},e_{\sigma^{(g)}(j)}),(\hat{s}^{(g)}_j,\hat{e}^{(g)}_j)\big)
  +(1-\alpha)\,\mathrm{Sim}\big(c_{\sigma^{(g)}(j)},\,\hat{c}^{(g)}_j\big), & \sigma^{(g)}(j)\neq 0,\\[2pt]
0, & \sigma^{(g)}(j)=0 .
\end{cases}
\end{equation}
where the final correctness reward is the average of $r^{(g)}_j$ over all the GT segments, i.e.\ $R^{(g)}_{\text{DAC}}=\frac{1}{N_g}\sum_{j=1}^{N^{(g)}_p} r^{(g)}_j$.

\subsection{Token-wise Fine-grained Advantage}
\label{subsec:adv}

Prior work on dense audio captioning \citep{shao2026listening_lat_audio} assigns a single scalar reward to the whole response. The direct analogue here mixes the three response-level rewards of \cref{eq:cag_agg,eq:ad,eq:dac_seg} with a format reward $R^{(g)}_{\text{fmt}}$, and normalizes the mixture group-wise as in standard GRPO:
\begin{equation}
\label{eq:seq_baseline}
R^{(g)}_{\text{seq}} = \lambda_{1} R^{(g)}_{\text{CAG}} + \lambda_{2} R^{(g)}_{\text{AD}}
+ \lambda_{3} R^{(g)}_{\text{DAC}} + \lambda_{4} R^{(g)}_{\text{fmt}},
\qquad
A^{(g)}_{\text{seq}} = \frac{R^{(g)}_{\text{seq}}-\mathrm{mean}_g(R_{\text{seq}})}
{\mathrm{std}_g(R_{\text{seq}})+\epsilon},
\end{equation}
assigning $\hat{A}^{(g)}_t = A^{(g)}_{\text{seq}}$ to every token. This provides a single coarse, sequence-level reward for the entire rollout. This issue is exacerbated in the case of long audios where the model response for captioning is inherently long leading to poor supervision. We report this variant as an ablation in \S\ref{sec:ablations}.

\textbf{Anchored per-segment advantages.} We propose more fine-grained advantage computation for each of $R_\text{CAG}$, $R_{\text{AD}}$ and $R_\text{DAC}$, instead of a sequence-level advantage in \cref{eq:seq_baseline}. Each reward is an average of per-segment scores (underbraced in \cref{eq:cag,eq:ad,eq:dac_seg}), so we normalize the segment-level rewards directly and let each token inherit the advantage of the segment it spells out. For GRPO advantage calculation, we define different anchors for the rewards as below.
 
\textbf{$R_\text{DAC}$: GT-anchored.} We normalize the segments across the GT-segments they are matched to according to the inverse assignment of $\sigma^{(g)}$ and assign a reward 0 from a rollout that missed that GT-segment. For GT event $i$,
let $\varphi_i(g)$ be the index of the segment in rollout $g$ matched to $i$, and define the per-rollout score $u^{(g)}_i$ as the segment-level reward $r^{(g)}_{\varphi_i(g)}$
so that $\{u^{(g)}_i\}_{g=1}^{G}$ always contains exactly $G$ entries. 
The segment advantage is then,
\begin{equation}
A^{\text{DAC},(g)}_i \;=\; \frac{u^{(g)}_i-\mu_i}{\varsigma_i+\epsilon},
\qquad
\mu_i=\frac{1}{G}\sum_g u^{(g)}_i,\quad
\varsigma_i=\mathrm{std}_g\big(u^{(g)}_i\big).
\end{equation}
This compares each rollout against the other rollouts \emph{on the same event}, so credit is not
diluted by how easy or hard the remaining events in the clip happen to be.
 
\textbf{$R_\text{CAG}$: window-anchored.} The counterfactual reference $\hat{D}'$ is decoded once and shared across the group, so the silenced windows $\mathcal{W}$ are fixed and independent of any individual rollout. For window $w$, we take the window score $q^{(g)}_w$ of \cref{eq:cag_agg} and normalize it within that window,
\begin{equation}
A^{\text{CAG},(g)}_w \;=\; \frac{q^{(g)}_w-\mu_w}{\varsigma_w+\epsilon},
\qquad
\mu_w=\tfrac{1}{G}\textstyle\sum_g q^{(g)}_w,\quad \varsigma_w=\mathrm{std}_g\big(q^{(g)}_w\big).
\end{equation}
Localizing the penalty this way means a rollout that hallucinates inside one silenced window is not down-weighted on the segments it predicts correctly elsewhere in the clip. 
 
\textbf{$R_{\text{AD}}$: group-pooled.} Attention density is defined on a rollout's own predicted span and admits no shared anchor across rollouts. We therefore pool the densities over all predicted segments of all $G$ rollouts, $\mathcal{S}=\{(g,j): 1\!\le\! g\!\le\! G,\ 1\!\le\! j\!\le\! N^{(g)}_p\}$, and standardize against that pool:
\begin{equation}
A^{\text{AD},(g)}_j \;=\; \frac{\rho^{(g)}_j-\mu_{\text{AD}}}{\varsigma_{\text{AD}}+\epsilon},
\quad
\mu_{\text{AD}}=\frac{1}{|\mathcal{S}|}\!\!\sum_{(g,j)\in\mathcal{S}}\!\!\rho^{(g)}_j,
\quad
\varsigma_{\text{AD}}=\mathrm{std}_{(g,j)\in\mathcal{S}}\big(\rho^{(g)}_j\big).
\end{equation}
Each segment is thus scored against the density the group achieves on average, so segments that attend to their own claimed time span more sharply than the group norm are reinforced individually rather than through a response-level average.
 
\textbf{Token assignment.} Every token inside the span $\mathcal{G}^{(g)}_j$ inherits the three
segment-level advantages of its segment; format tokens receive only the
sequence-level term $A^{(g)}_{\text{seq}}$, now obtained from \cref{eq:seq_baseline} with $\lambda_{1}\!=\!\lambda_{2}\!=\!\lambda_{3}\!=\!0$, i.e.\ by group-normalizing $R^{(g)}_{\text{fmt}}$ alone:
\begin{equation}
\label{eq:tok_assign}
\hat{A}^{(g)}_t \;=\; A^{(g)}_{\text{seq}}
\;+\;\sum_{j=1}^{N^{(g)}_p} \ind \big[t\in\mathcal{G}^{(g)}_j\big]
\Big(\eta_{1} A^{\text{CAG},(g)}_{w(j)}
\;+\;\eta_{2} A^{\text{AD},(g)}_{j}
\;+\;\eta_{3} A^{\text{DAC},(g)}_{\sigma^{(g)}(j)}\Big),
\end{equation}
where $w(j)\in\mathcal{W}$ is the silenced window containing segment $j$. The weights $\eta_{1},\eta_{2},\eta_{3}$ are hyperparameters to balance the three signals. 
Finally, ATTEST optimizes the clipped GRPO objective with these token-wise advantages:
\begin{equation}
\small
\mathcal{J}(\theta)=\mathbb{E}\Big[\tfrac{1}{G}\!\sum_{g=1}^{G}\tfrac{1}{|o^{(g)}|}\!
\sum_{t=1}^{|o^{(g)}|}\!\min\!\big(\varrho^{(g)}_t \hat{A}^{(g)}_t,\;
\mathrm{clip}(\varrho^{(g)}_t,1{-}\varepsilon_{\text{clip}},1{+}\varepsilon_{\text{clip}})\hat{A}^{(g)}_t\big)\Big]
-\;\kappa\,\mathbb{D}_{\mathrm{KL}}\!\left[\pi_\theta\,\|\,\pi_{\mathrm{ref}}\right],
\end{equation}
where $\varrho^{(g)}_t=\pi_\theta(o^{(g)}_t\mid x,o^{(g)}_{<t})/\pi_{\theta_{\text{old}}}(o^{(g)}_t\mid x,o^{(g)}_{<t})$.

%% file: secs/5_experiments.tex
\begin{table}[]
\centering
\caption{Dense audio captioning results on EgoSCAPE and AudioSet-SL datasets comparing ATTEST to other audio and omni LLM baselines. EvtF1@0.5 and SegF1 are micro. Hal\% and SrcF1@0.5 are macro. * denotes sliding-window evaluation described in \S\ref{appsec:baseline_details}. TimeAudio and AudioFlamingo 2 did not adhere to the instructions to produce segment-wise sound sources so we have omitted SrcF1 for them.}
\label{tab:dense_cap_results}
\vspace{-1em}
\resizebox{\columnwidth}{!}{%
\begin{tabular}{l|ccccccc|cccccc}
\hline \hline
\multicolumn{1}{c|}{\multirow{2}{*}{\textbf{Model}}} & \multicolumn{7}{c|}{\textbf{EgoSCAPE Dense Audio Captioning}} & \multicolumn{6}{c}{\textbf{AudioSet-SL}} \\ \cline{2-14} 
\multicolumn{1}{c|}{} & \textbf{F@0.3} & \textbf{F@0.5} & \textbf{F@0.7} & \textbf{EvtF1} & \textbf{SegF1} & \textbf{Hal\%$\downarrow$} & \textbf{SrcF1} & \textbf{F@0.3} & \textbf{F@0.5} & \textbf{F@0.7} & \textbf{EvtF1} & \textbf{SegF1} & \textbf{Hal\%$\downarrow$} \\ \hline
\multicolumn{14}{c}{\emph{Proprietary Models}} \\ \hline
Gemini 3.1 Pro & 0.235 & 0.213 & 0.201 & 0.084 & 0.276 & 31.1 & 0.371 & 0.207 & 0.192 & 0.181 & 0.079 & 0.126 & 22.7 \\
Gemini 3.6 Flash & 0.242 & 0.215 & 0.200 & 0.091 & 0.197 & 38.9 & 0.351 & 0.217 & 0.200 & 0.172 & 0.037 & 0.066 & 30.3 \\
GPT 4o Audio & 0.226 & 0.208 & 0.198 & 0.059 & 0.186 & 48.4 & 0.359 & 0.208 & 0.184 & 0.167 & 0.051 & 0.058 & 50.9 \\\hline
\multicolumn{14}{c}{\emph{Captioning Models}} \\ \hline
AF-Next Cap. & 0.244 & 0.218 & 0.201 & 0.099 & 0.147 & 52.7 & 0.325 & 0.239 & 0.207 & 0.184 & 0.058 & 0.087 & 44.6 \\
Qwen3 Omni Cap. & 0.270 & 0.232 & 0.206 & 0.257 & 0.335 & 64.4 & 0.369 & 0.257 & 0.220 & 0.190 & 0.048 & 0.057 & 63.6 \\ \hline 
\multicolumn{14}{c}{\emph{Audio and Omni LLMs}} \\ \hline
Qwen2 Audio* & 0.207 & 0.199 & 0.196 & 0.048 & 0.127 & 51.0 & 0.232 & 0.207 & 0.191 & 0.182 & 0.054 & 0.076 & 43.1 \\
Kimi-Audio* & 0.188 & 0.183 & 0.177 & 0.042 & 0.034 & 42.9 & 0.207 & 0.224 & 0.198 & 0.175 & 0.061 & 0.107 & 33.9 \\
TimeAudio* & 0.340 & 0.310 & 0.271 & 0.256 & 0.344 & 35.2 & - & 0.426 & 0.389 & 0.352 & 0.353 & 0.372 & 36.3 \\
AudioFlamingo 2 & 0.193 & 0.193 & 0.192 & 0.004 & 0.001 & 59.7 & - & 0.192 & 0.176 & 0.167 & 0.079 & 0.084 & 55.5 \\
AF-Next & 0.181 & 0.178 & 0.176 & 0.022 & 0.014 & 45.9 & 0.030 & 0.226 & 0.202 & 0.184 & 0.089 & 0.114 & 48.5 \\ \hline
Qwen2.5 Omni & 0.178 & 0.177 & 0.176 & 0.021 & 0.035 & 48.2 & 0.165 & 0.220 & 0.202 & 0.192 & 0.124 & 0.190 & 20.5 \\
+ SFT & 0.412 & 0.367 & 0.313 & 0.320 & 0.303 & 47.2 & 0.564 & 0.448 & 0.403 & 0.362 & 0.336 & 0.337 & 25.8 \\
+ TimePro-RL & 0.445 & 0.410 & 0.358 & 0.349 & 0.348 & 31.3 & 0.642 & 0.475 & 0.430 & 0.385 & 0.347 & 0.370 & 20.1 \\
\rowcolor{metalight}
\textbf{+ ATTEST} & \textbf{0.461} & \textbf{0.427} & \textbf{0.372} & \textbf{0.371} & \textbf{0.355} & \textbf{30.5} & \textbf{0.656} & \textbf{0.489} & \textbf{0.442} & \textbf{0.399} & \textbf{0.354} & 0.389 & \textbf{18.2} \\ \hline
AudioFlamingo 3 & 0.196 & 0.194 & 0.193 & 0.042 & 0.010 & 40.3 & - & 0.204 & 0.189 & 0.177 & 0.119 & 0.131 & 32.4 \\
+ SFT & 0.376 & 0.344 & 0.298 & 0.258 & 0.248 & 44.4 & 0.587 & 0.453 & 0.406 & 0.362 & 0.345 & 0.396 & 36.3 \\
+ TimePro-RL & 0.421 & 0.388 & 0.328 & 0.335 & 0.329 & 34.0 & 0.620 & 0.489 & 0.440 & 0.384 & 0.369 & 0.385 & 29.2 \\
\rowcolor{metalight}
\textbf{+ ATTEST} & \textbf{0.437} & \textbf{0.400} & \textbf{0.342} & \textbf{0.346} & \textbf{0.345} & \textbf{31.8 }& \textbf{0.641} & \textbf{0.503} & \textbf{0.451} & \textbf{0.398} & \textbf{0.380} & \textbf{0.400} & \textbf{25.3 }\\ 
\hline \hline
\end{tabular}%
}
\end{table}

\section{Experiments}
\label{sec:experiments}


\textbf{Setup.} We use Qwen2.5 Omni \citep{xu2025qwen25omnitechnicalreport} and AudioFlamingo 3 \citep{ghosh2026audioflamingo3_af3} as base models (with Qwen3 Omni \citep{xu2025qwen3omni} in the Appendix). RL training uses the AudioSet-SL training set \citep{shawn2021audioset_sl}, supplemented with long audio generated via the pipeline in \S\ref{sec:egoscape_data} from Ego4D videos disjoint from EgoSCAPE. We set $(\eta_1, \eta_2, \eta_3) = (1, 0.2, 1)$ and $\lambda_4 = 1$; the sequence-level ablation reuses these weights as $(\lambda_1, \lambda_2, \lambda_3)$. We apply GRPO with a rollout size of 8 and an effective batch size of 16, using LoRA finetuning at a learning rate of $5 \times 10^{-6}$ (details in \S\ref{appsubsec:training_details}).

\textbf{Evaluation.} We evaluate on temporal grounding and general audio understanding benchmarks. For dense audio captioning, we report results on the AudioSet-SL test set \citep{shawn2021audioset_sl} following TimeAudio \citep{wang2026listening_timeaudio_ftar}, alongside EgoSCAPE. For temporal audio grounding, we report zero-shot results on Clotho-Moment \citep{munakata2025language_clotho_moment_lamr}, UnAV-100 \citep{geng2023dense_unav100} subset from \cite{munakata2025language_clotho_moment_lamr} and CASTELLA \citep{munakata2026castella}. For general audio understanding, we use MMAR \citep{ma2026mmar} \& LongAudioBench \citep{ghosh2025audioflamingo2_af2}.

\textbf{Metrics.} For dense captioning and audio timeline generation, we report FENSE \citep{zhou2022can_fense} scores at IoU thresholds of 0.3, 0.5, and 0.7 following \cite{shao2026listening_lat_audio}, along with Event-F1 and Segment-F1 \citep{mesaros2016metrics_sed}. Following TAC \citep{kumar2026tac}, we use FLAM \citep{flam2025} to measure hallucination rates over predicted segments unmatched to any ground truth: a segment is deemed hallucinated if its \emph{confidence}, the maximum FLAM similarity between its audio frames and its caption, falls below 0.15. For EgoSCAPE, we additionally report sound source detection F1. For temporal audio grounding, we report recall at IoU thresholds of 0.5 and 0.7 and mean IoU (more in \S\ref{appsubsec:evaluation_metrics}).

\textbf{Baselines.} We compare against Gemini 3.1 Pro \citep{gemini31pro2026}, Gemini 3.6 Flash \citep{gemini36flash}, GPT 4o Audio \citep{openai2024gpt4ocard}, Qwen2 Audio \citep{chu2024qwen2audio}, Kimi-Audio \citep{ding2025kimiaudio}, TimeAudio \citep{wang2026listening_timeaudio_ftar}, AudioFlamingo 2, 3 \& Next \citep{ghosh2025audioflamingo2_af2,ghosh2026audioflamingo3_af3,ghosh2026audioflamingonext_afnext}, Qwen2.5 Omni \citep{xu2025qwen25omnitechnicalreport}, and Qwen3 Omni \citep{xu2025qwen3omni}. As Qwen2 Audio, Kimi-Audio, and TimeAudio cannot process 5-minute audio, we use a sliding-window approach. We also compare with TimePro-RL \citep{shi2026towards_timepro_rl} framework adapted for our setup (details in \S\ref{appsec:baseline_details}).

\subsection{Audio Captioning}
\label{subsec:results_captioning}

\cref{tab:dense_cap_results} (extended in \S\ref{appsubsec:audio_captioning_results} \cref{apptab:dense_cap_results_extended}) reports DAC results on EgoSCAPE and AudioSet-SL \citep{wang2026listening_timeaudio_ftar}. On both datasets, ATTEST achieves the best DAC performance, substantially improving over its base models while also markedly reducing hallucinations. Dedicated captioning models outperform most off-the-shelf audio and omni LLMs but hallucinate the most, whereas proprietary models hallucinate far less, with Gemini 3.1 Pro achieving the lowest hallucination rate among baselines on EgoSCAPE. Audio timeline generation results (\S\ref{appsubsec:audio_captioning_results} \cref{tab:egoscape_timeline_results}) follow similar trends.

\begin{table}[]
\centering
\caption{Zero-shot temporal audio grounding results on Clotho-Moment, UnAV-100 and CASTELLA datasets comparing ATTEST to other audio and omni LLM baselines. }
\vspace{-1em}
\label{tab:tag_results_zeroshot}
\resizebox{0.9\columnwidth}{!}{%
\begin{tabular}{l|ccc|ccc|ccc}
\hline \hline
\multicolumn{1}{c|}{\multirow{2}{*}{\textbf{Model}}} & \multicolumn{3}{c|}{\textbf{Clotho-Moment}} & \multicolumn{3}{c|}{\textbf{UnAV-100}} & \multicolumn{3}{c}{\textbf{CASTELLA}} \\ \cline{2-10} 
\multicolumn{1}{c|}{} & \textbf{R@0.5} & \textbf{R@0.7} & \textbf{mIoU} & \textbf{R@0.5} & \textbf{R@0.7} & \textbf{mIoU} & \textbf{R@0.5} & \textbf{R@0.7} & \textbf{mIoU} \\ \hline
\multicolumn{10}{c}{\emph{Proprietary Models}} \\ \hline
Gemini 3.1 Pro & 0.385 & 0.296 & 0.344 & 0.340 & 0.240 & 0.294 & 0.080 & 0.041 & 0.081 \\
Gemini 3.6 Flash & 0.581 & 0.431 & 0.536 & 0.570 & 0.410 & 0.524 & 0.233 & 0.124 & 0.223 \\
GPT 4o Audio & 0.106 & 0.031 & 0.185 & 0.200 & 0.060 & 0.216 & 0.042 & 0.018 & 0.062 \\ \hline
\multicolumn{10}{c}{\emph{Finetuned Temporal Grounding Models}} \\ \hline
AM-DETR & 0.882 & 0.827 & 0.809 & 0.580 & 0.410 & 0.428 & 0.094 & 0.062 & 0.103 \\ 
SpotSound-A & 0.908 & 0.853 & 0.868 & 0.780 & 0.620 & 0.725 & 0.180 & 0.115 & 0.181 \\
UVCOM & 0.885 & 0.832 & 0.820 & 0.590 & 0.360 & 0.518 & \underline{0.302} & \underline{0.191} & \underline{0.317} \\ \hline
\multicolumn{10}{c}{\emph{Audio and Omni LLMs}} \\ \hline
Qwen2 Audio* & 0.043 & 0.005 & 0.131 & 0.080 & 0.040 & 0.181 & 0.053 & 0.020 & 0.097 \\
Kimi-Audio* & 0.088 & 0.025 & 0.164 & 0.180 & 0.030 & 0.265 & 0.080 & 0.022 & 0.146 \\
AudioFlamingo 2 & 0.009 & 0.003 & 0.023 & 0.040 & 0.010 & 0.056 & 0.011 & 0.005 & 0.016 \\
AF-Next & 0.004 & 0.002 & 0.019 & 0.030 & 0.030 & 0.069 & 0.008 & 0.002 & 0.013 \\ \hline
Qwen2.5 Omni & 0.409 & 0.141 & 0.426 & 0.470 & 0.280 & 0.458 & 0.092 & 0.048 & 0.116 \\
+ SFT & 0.875 & 0.821 & 0.830 & 0.790 & 0.600 & 0.693 & 0.224 & 0.123 & 0.269 \\
+ TimePro-RL & 0.895 & 0.844 & 0.859 & 0.800 & 0.620 & 0.711 & 0.252 & 0.143 & 0.295 \\
\rowcolor{metalight}
\textbf{+ ATTEST} & \textbf{0.909} & \textbf{0.859} & \textbf{0.876} & \textbf{0.820} & \textbf{0.650} & \textbf{0.729} & \textbf{0.297} & \textbf{0.186} & \textbf{0.312} \\ \hline
AudioFlamingo 3 & 0.142 & 0.063 & 0.163 & 0.120 & 0.030 & 0.147 & 0.008 & 0.004 & 0.013 \\
+ SFT & 0.850 & 0.754 & 0.801 & 0.780 & 0.620 & 0.708 & 0.147 & 0.095 & 0.170 \\
+ TimePro-RL & 0.892 & 0.799 & 0.836 & 0.810 & 0.650 & 0.744 & 0.168 & 0.110 & 0.189 \\
\rowcolor{metalight}
\textbf{+ ATTEST} & \textbf{0.903} & \textbf{0.819} & \textbf{0.854} & \textbf{0.850} & \textbf{0.660} & \textbf{0.753} & \textbf{0.202} & \textbf{0.123} & \textbf{0.266} \\ 
\hline \hline
\end{tabular}%
}
\vspace{-1em}
\end{table}

\subsection{Temporal Audio Grounding}
\label{subsec:results_grounding}

\cref{tab:tag_results_zeroshot} (extended in \S\ref{appsubsec:tag_results_extended}  \cref{apptab:tag_results_zeroshot_expanded}) reports zero-shot TAG results on Clotho-Moment, UnAV-100, and CASTELLA. ATTEST substantially improves over its base models and outperforms all LLM-based baselines. Despite never seeing these benchmarks' training sets, it matches or surpasses dedicated grounding methods fine-tuned on them \citep{munakata2025language_clotho_moment_lamr,munakata2026castella,sun2026spotsound}, demonstrating strong generalization. Results on EgoSCAPE TAG samples (\S\ref{appsubsec:tag_results_extended} \cref{tab:tag_results_egoscape}) show similar trends.

\begin{table}[]
\centering
\caption{Ablation study to show the effect of different techniques and rewards used in ATTEST framework. TA: Token-wise fine-grained advantage.}
\label{tab:ablation_reward}
\vspace{-1em}
\resizebox{\columnwidth}{!}{%
\begin{tabular}{l|cccc|cccc|cc}
\hline \hline
\multicolumn{1}{c|}{\multirow{3}{*}{\textbf{Model}}} & \multirow{3}{*}{\textbf{TA}} & \multirow{3}{*}{\textbf{$R_\text{CAG}$}} & \multirow{3}{*}{\textbf{$R_\text{AD}$}} & \multirow{3}{*}{\textbf{$R_\text{DAC}$}} & \multicolumn{4}{c|}{\textbf{EgoSCAPE Dense Captioning}} & \multicolumn{2}{c}{\textbf{Temporal Grounding}} \\ \cline{6-11}
\multicolumn{1}{c|}{} &  &  &  &  & \multicolumn{2}{c}{\textbf{Overall}} & \multicolumn{2}{c|}{\textbf{Long (\textgreater{}120s)}} & \textbf{Clotho-Moment} & \textbf{CASTELLA} \\ \cline{6-11}
\multicolumn{1}{c|}{} &  &  &  &  & \textbf{F@0.5} & \textbf{Hal\%$\downarrow$} & \textbf{F@0.5} & \textbf{Hal\%$\downarrow$} & \textbf{R@0.5} & \textbf{R@0.5} \\ \hline
Qwen2.5 Omni & - & - & - & - & 0.177 & 48.2 & 0.169 & 55.2 & 0.409 & 0.092 \\
+ SFT & - & - & - & - & 0.367 & 47.2 & 0.321 & 50.1 & 0.875 & 0.224 \\
+ Vanilla GRPO & \xmark & \xmark & \xmark & \cmark & 0.395 & 38.2 & 0.379 & 41.2 & 0.879 & 0.236 \\
\rowcolor{metalight}
\textbf{+ ATTEST} & \cmark & \cmark & \cmark & \cmark & \textbf{0.427} & \textbf{30.5} & \textbf{0.414} & \textbf{34.2} & \textbf{0.909} & \textbf{0.297} \\
w/o $R_\text{CAG}$ & \cmark & \xmark & \cmark & \cmark & 0.405 & 37.2 & 0.382 & 38.4 & 0.898 & 0.252 \\
w/o $R_\text{AD}$ & \cmark & \cmark & \xmark & \cmark & 0.409 & 33.7 & 0.385 & 35.8 & 0.896 & 0.260 \\
w/o $R_\text{DAC}$ & \cmark & \cmark & \cmark & \xmark & 0.382 & 34.1 & 0.375 & 36.5 & 0.881 & 0.241 \\
w/o Token-adv. & \xmark & \cmark & \cmark & \cmark & 0.411 & 32.1 & 0.398 & 34.9 & 0.901 & 0.268 \\ \hline \hline
\end{tabular}%
}
\vspace{-1em}
\end{table}

\subsection{Ablations}
\label{sec:ablations}
\noindent \textbf{Role of individual rewards.} \cref{tab:ablation_reward} ablates the components of ATTEST (\S\ref{sec:attest_method}), reporting FENSE at tIoU 0.5 and hallucination rates on all EgoSCAPE clips and on long clips ($>$120s). Removing $R_{\text{AD}}$ or $R_{\text{CAG}}$ substantially increases hallucinations and lowers FENSE. Ablating $R_{\text{DAC}}$, which jointly optimizes text similarity and tIoU, causes the largest FENSE drop, while replacing token-wise with sequence advantage degrades performance across all metrics.

\noindent \textbf{Hyper-parameter and reward design.} Effects of the reward weights ($\eta_1$, $\eta_2$, $\eta_3$), the corruption operator in $R_{\text{CAG}}$, and the attention layer in $R_{\text{AD}}$ are analyzed in \S\ref{subsubsec:hyperparam_reward_design}.

\subsection{General audio benchmarks}
\label{subsec:results_general_benchmarks}
\vspace{-0.5em}
\begin{wraptable}{r}{0.55\textwidth}
\centering
\vspace{-3em}
\caption{LongAudioBench and MMAR results.}
\vspace{-1em}
\label{tab:general_audio_benchmarks}
\resizebox{\linewidth}{!}{%
\begin{tabular}{l|cc|cc}
\hline\hline
\multicolumn{1}{c|}{\multirow{2}{*}{\textbf{Model}}} & \multicolumn{2}{c|}{\textbf{MMAR}} & \multicolumn{2}{c}{\textbf{LongAudioBench}} \\ \cline{2-5} 
\multicolumn{1}{c|}{} & \textbf{All} & \textbf{Temporal} & \textbf{All} & \textbf{Temporal} \\ \hline
Qwen2.5 Omni & 63.2 & 56.0 & 61.5 & 60.1 \\
+ TimePro-RL & 61.0 & 55.4 & 58.3 & 59.5 \\
\rowcolor{metalight}
+ \textbf{ATTEST} & \textbf{64.4} & \textbf{58.2} & \textbf{62.8} & \textbf{62.2} \\
+ ATTEST w/ markers & 62.2 & 56.9 & 60.5 & 61.2 \\ \hline
AudioFlamingo 3 & 60.4 & 53.2 & 57.7 & 53.9 \\
+ TimePro-RL & 58.3 & 51.0 & 55.2 & 52.5 \\
\rowcolor{metalight}
+ \textbf{ATTEST} & \textbf{62.1} & \textbf{57.9} & \textbf{59.6} & \textbf{57.2} \\
+ ATTEST w/ markers & 59.2 & 52.8 & 56.9 & 53.5 \\\hline\hline
\end{tabular}%
}
\vspace{-1em}
\end{wraptable}
\cref{tab:general_audio_benchmarks} (extended in \cref{apptab:general_audio_benchmarks_expanded}) reports results on MMAR and LongAudioBench. As neither benchmark tags temporal samples explicitly, we use an LLM to identify a temporal subset and report it separately. ATTEST improves overall performance, with larger gains on the temporal subset, indicating that its temporal improvements transfer to general audio understanding. In contrast, TimePro-RL degrades performance under identical training data and base models, and so does an ATTEST variant with TimePro-RL-style temporal markers, suggesting that temporal markers hurt original audio understanding capabilities of different models.

\subsection{Analysis}
\vspace{-0.5em}
\begin{wrapfigure}{r}{0.5\textwidth}
    \vspace{-4em}
    \centering
    \includegraphics[width=0.5\textwidth]{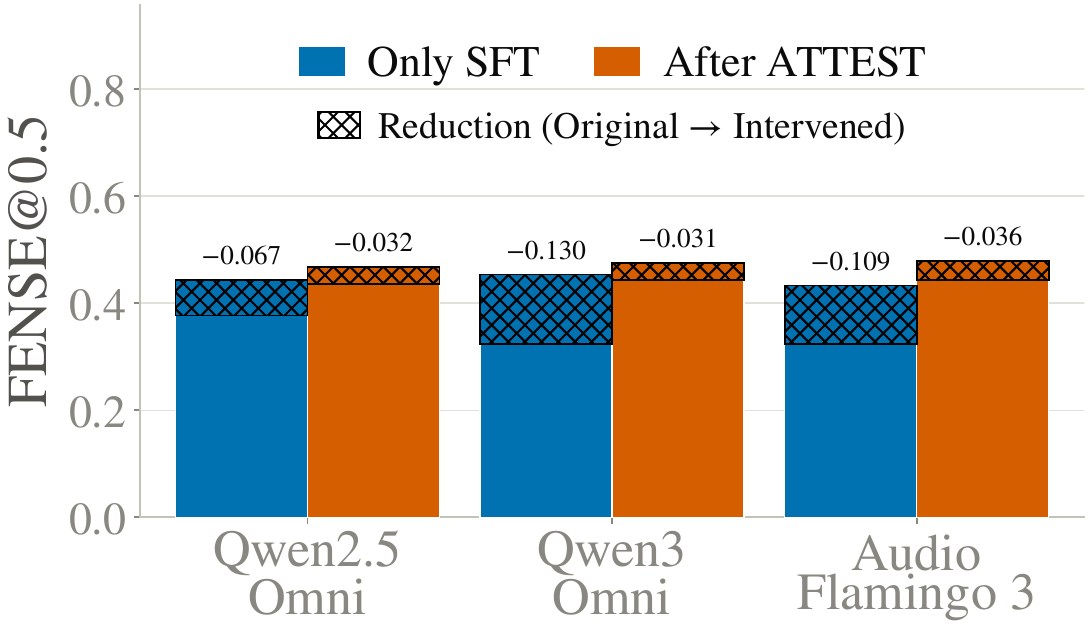}
    \caption{Perf. drop on intervened clips.}
    \label{fig:robustness_intervention}
    \vspace{-1em}
\end{wrapfigure}
\textbf{Robustness to hallucination.} To test robustness to audio intervention, we randomly replace segments of EgoSCAPE audio with AudioSet-SL clips and compute FENSE@0.5 on the replaced segments (\cref{fig:robustness_intervention}). ATTEST-tuned models accurately predict sound events and boundaries even for such out-of-context segments, whereas SFT-tuned models degrade significantly. Similar analysis on silence intervention is present in \S\ref{appsubsec:robustness_halluc_extended}.

\vspace{-0.5em}
\textbf{Attention analysis and importance of real data.} In \S\ref{subsec:attention_behavior}, we show how attention patterns of the model differ after training with ATTEST. Additionally, we show the importance of real long data over synthetic data augmentation in \S\ref{appsubsec:real_data_importance}.

%% file: secs/6_conclusion.tex
\section{Conclusion}

We introduced EgoSCAPE, a long-audio benchmark for evaluating temporal grounding and captioning in LALMs, and ATTEST, an RL framework for improving their temporal grounding. ATTEST grounds responses in audio evidence via a counterfactual reward, steers attention toward the audio tokens each prediction describes, and provides fine-grained credit assignment through token-wise advantages. We believe this work marks a significant step toward stronger temporal grounding in LALMs, enabling better audio understanding and reasoning in general scenarios.

\subsection*{AI Use Statement}

In this work, we used generative AI tools for generating synthetic data sets, support qualitative and thematic data analysis, and assisting with cleaning and reformatting the dataset.
We have not used generative AI tools for developing theoretical models or conceptual frameworks, implementing methods, proposing or refining hypotheses, designing or providing feedback on research methodology or experiment, or for interpreting the results.
Formulating mathematical claims, providing critical ingredients for proving mathematical claims and assisting in writing of proofs are not applicable to this work.

Additionally, we have used generative AI tools for summarizing and analysing existing literature. The generated literature survey was manually verified for the correctness of citations and relevance to this work. We have also used AI tools for checking the correctness of grammar and language of this research paper.   




\subsection*{Ethics Statement}

EgoSCAPE is built on Ego4D~\citep{grauman2022ego4d}, which was collected with informed consent from camera wearers and under the ethics review of its contributing institutions. We use it under the Ego4D license and release only our annotations with clip identifiers, not the underlying audio or video. Our annotations can not identify individual speakers. We acknowledge that improved temporal audio grounding could be misused for audio surveillance; our models are trained for sound-event captioning and grounding rather than speaker identification, and their outputs should not be used in high-stakes settings without human review.

\subsection*{Reproducibility Statement}

We use open-weight models to construct EgoSCAPE so that the benchmark creation pipeline remains reproducible. All experimental details are provided in \S\ref{sec:experiments} and \S\ref{appsec:experimental_details}. Moreover, we demonstrate the efficacy of the proposed ATTEST method on open-source and open-weight models to ensure ease of access. All prompts used throughout this work are provided in \S\ref{appsec:prompts}. 

%% file: secs/A_appendix.tex
\clearpage
\appendix

{
        \centering
        \huge
        \vspace{0.5em} \textsc{Appendix} \\
        \vspace{1.0em}
}


{\noindent \Large \textsc{Table of Contents}
}

\begin{itemize}
    \item {\textsc{Limitations And Future Work}} \dotfill \ref{appsec:limitations}
    \item {\textsc{EgoSCAPE Benchmark Details}} \dotfill \ref{appsec:benchmark_details}
    \begin{itemize}
        \item {\small \textsc{Comparison with existing benchmarks}} \dotfill \ref{appsubsec:benchmark_comparison}
        \item {\small \textsc{Benchmark statistics}} \dotfill \ref{appsubec:benchmark_stats}
        \item {\small \textsc{Human verification details}} \dotfill \ref{appsubsec:human_verification_details}
    \end{itemize}
    \item {\textsc{Experimental Details}} \dotfill \ref{appsec:experimental_details}
    \begin{itemize}
        \item {\small \textsc{Training details}} \dotfill \ref{appsubsec:training_details}
        \item {\small \textsc{Evaluation metrics}} \dotfill \ref{appsubsec:evaluation_metrics}
        \item {\small \textsc{Baseline implementation details}} \dotfill \ref{appsec:baseline_details}
        \item {\small \textsc{Figure 3 details}} \dotfill \ref{appsubsec:figure3_details}
    \end{itemize}
    \item {\textsc{Extended Results}} \dotfill \ref{appsec:extended_results}
    \begin{itemize}
        \item {\small \textsc{Audio captioning}} \dotfill \ref{appsubsec:audio_captioning_results}
        \item {\small \textsc{Temporal audio grounding}} \dotfill \ref{appsubsec:tag_results_extended}
        \item {\small \textsc{General audio benchmarks}} \dotfill \ref{appsubsec:general_audio_benchmark_results}
        \item {\small \textsc{Hyper-parameter and reward design}} \dotfill \ref{subsubsec:hyperparam_reward_design}
        \item {\small \textsc{Robustness to hallucination}} \dotfill \ref{appsubsec:robustness_halluc_extended}
        \item {\small \textsc{Attention behavior}} \dotfill \ref{subsec:attention_behavior}
        \item {\small \textsc{Importance of real data}} \dotfill \ref{appsubsec:real_data_importance}
    \end{itemize}
    \item {\textsc{Prompts}} \dotfill \ref{appsec:prompts}
    
\end{itemize}

\section{Limitations And Future Work}
\label{appsec:limitations}

Our work represents a step forward in improving the temporal grounding and understanding of audio models. Although the EgoSCAPE benchmark evaluates how well current models ground events in long audio, there remains a need for benchmarks that probe deeper temporal reasoning, where models must use grounding information to solve multi-hop questions over long recordings. Furthermore, while we have so far evaluated the ATTEST framework on improving grounding in audio models, the same techniques could be extended to vision and omni-modal models to strengthen their temporal capabilities.

\section{EgoSCAPE Benchmark Details}
\label{appsec:benchmark_details}

\subsection{Comparison with existing benchmarks}
\label{appsubsec:benchmark_comparison}

\begin{table}[h]
\centering
\caption{Comparison of EgoSCAPE with existing temporal audio grounding benchmarks.}
\label{tab:benchmark_comparison}
\resizebox{\columnwidth}{!}{%
\begin{tabular}{l|ccccc|cccc}
\hline \hline
\multicolumn{1}{c|}{\textbf{Dataset}} & \textbf{Type} & \textbf{\begin{tabular}[c]{@{}c@{}}\# unique\\ audios\end{tabular}} & \textbf{\# samples} & \textbf{\begin{tabular}[c]{@{}c@{}}Avg.\\ Dur.\end{tabular}} & \textbf{\begin{tabular}[c]{@{}c@{}}Max.\\ Dur.\end{tabular}} & \textbf{\begin{tabular}[c]{@{}c@{}}Dense\\ Captions\end{tabular}} & \textbf{\begin{tabular}[c]{@{}c@{}}Source\\ Annot.\end{tabular}} & \textbf{\begin{tabular}[c]{@{}c@{}}Audio\\ Timeline\end{tabular}} & \textbf{\begin{tabular}[c]{@{}c@{}}Temporal\\ Grounding\end{tabular}} \\ \hline
FTAR-test & Real & 1224 & 1738 & 9.9 & 10.0 & \cmark & \xmark & \xmark & \cmark \\
LAT-Bench & Real & 247 & 2722 & 600.0 & 1790.2 & \xmark & \xmark & \cmark & \cmark \\
Clotho-Moment & Synthetic & 7569 & 7569 & 60.0 & 60.0 & \xmark & \xmark & \xmark & \cmark \\
SpotSound-Bench & Synthetic & 300 & 300 & 52.9 & 60.0 & \xmark & \xmark & \xmark & \cmark \\
CASTELLA & Real & 640 & 1302 & 206.0 & 300.0 & \xmark & \xmark & \xmark & \cmark \\ \hline
\textbf{EgoSCAPE} & \textbf{Real} & \textbf{700} & \textbf{4200} & \textbf{67.6} & \textbf{300.0} & \cmark & \cmark & \cmark & \cmark \\ \hline \hline
\end{tabular}%
}
\end{table}

\cref{tab:benchmark_comparison} shows the comparison of our benchmark with existing benchmarks which are not rich in annotations especially for dense annotations over long audio clips \citep{wang2026listening_timeaudio_ftar,shao2026listening_lat_audio,sun2026spotsound,munakata2025language_clotho_moment_lamr,munakata2026castella}.

\subsection{Benchmark Statistics}
\label{appsubec:benchmark_stats}

\begin{figure}[]
    \centering
    \begin{subfigure}[b]{0.49\textwidth}
        \includegraphics[width=\linewidth]{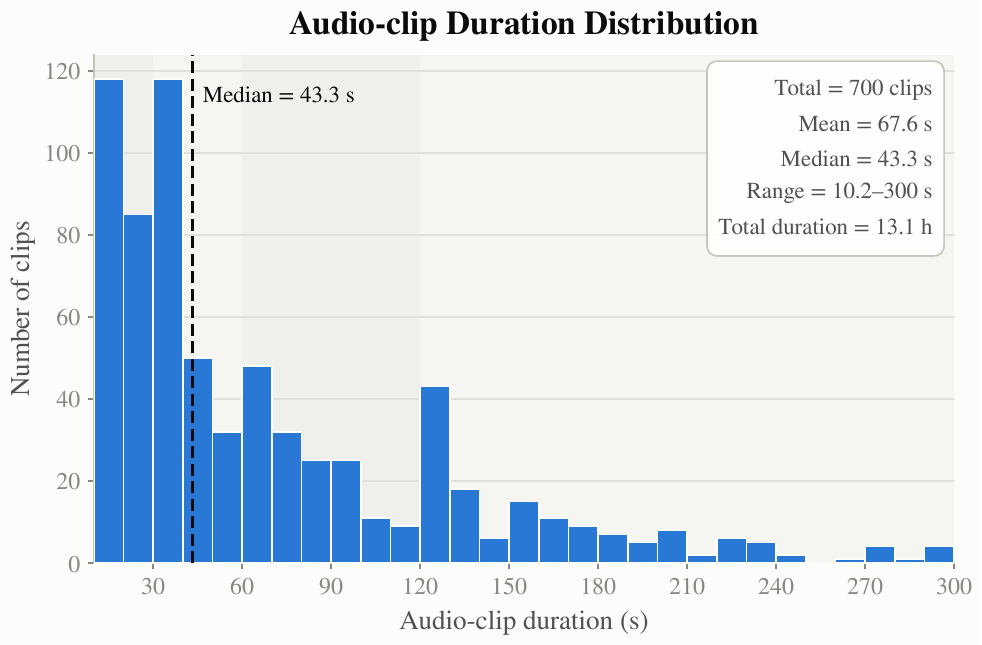}
        \caption{Audio-clip duration distribution.}
        \label{subfig:egoscape_duration_dist}
    \end{subfigure}
    \hfill
    \begin{subfigure}[b]{0.49\textwidth}
        \includegraphics[width=\linewidth]{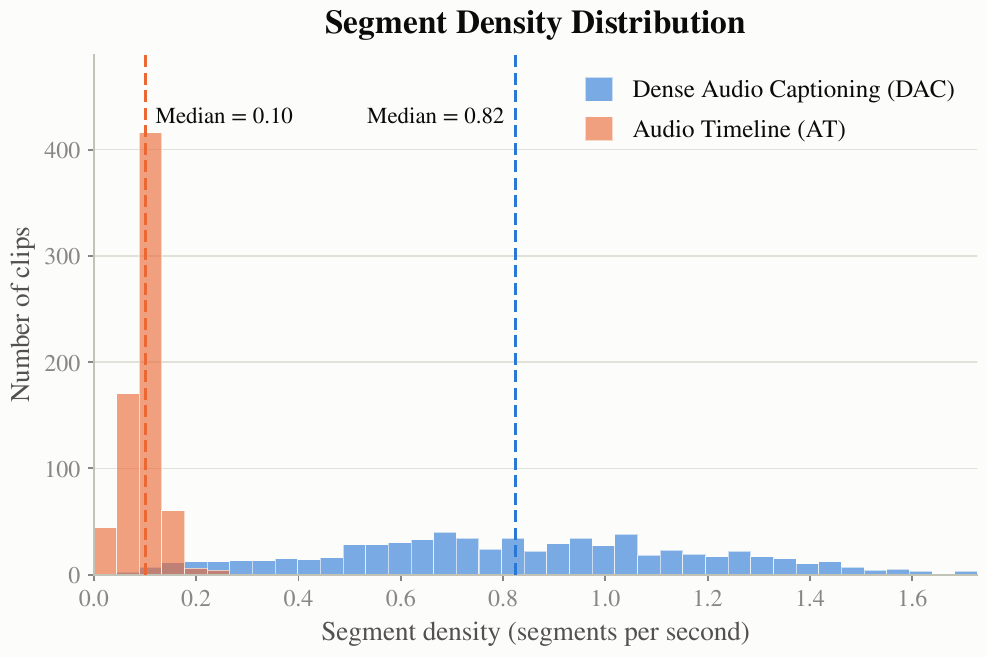}
        \caption{Segment density distribution.}
        \label{subfig:egoscape_seg_density_dist}
    \end{subfigure}
    
    \vspace{0.5em}
    \begin{subfigure}[b]{0.49\textwidth}
        \includegraphics[width=\linewidth]{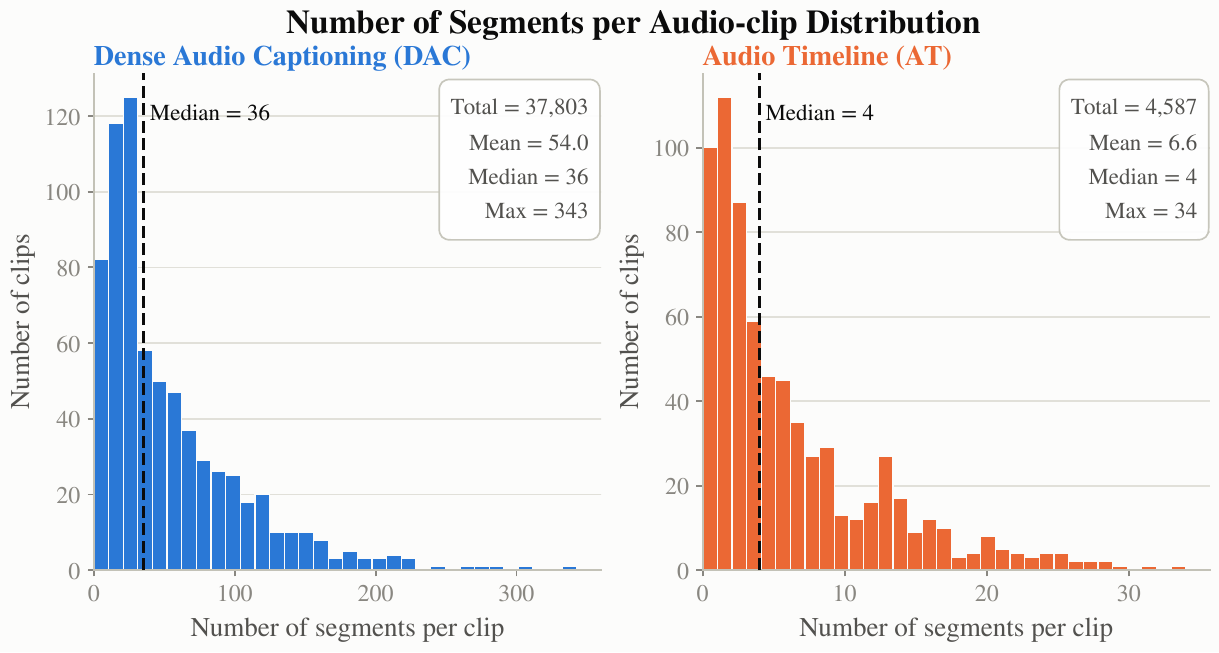}
        \caption{Number of segments per audio-clip.}
        \label{subfig:egoscape_seg_per_audio_dist}
    \end{subfigure}
    \hfill
    \begin{subfigure}[b]{0.49\textwidth}
        \includegraphics[width=\linewidth]{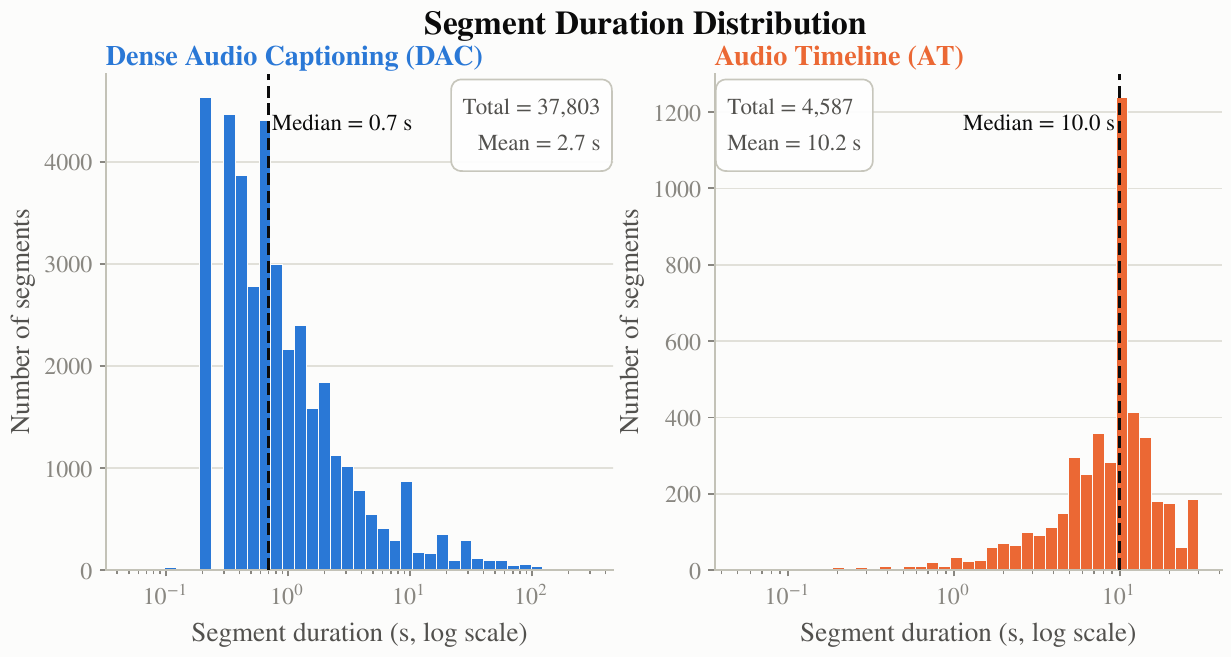}
        \caption{Segment duration distribution.}
        \label{subfig:egoscape_seg_duration_dist}
    \end{subfigure}

    \vspace{0.5em}
    
    \begin{subfigure}[b]{0.55\textwidth}
        \includegraphics[width=\linewidth]{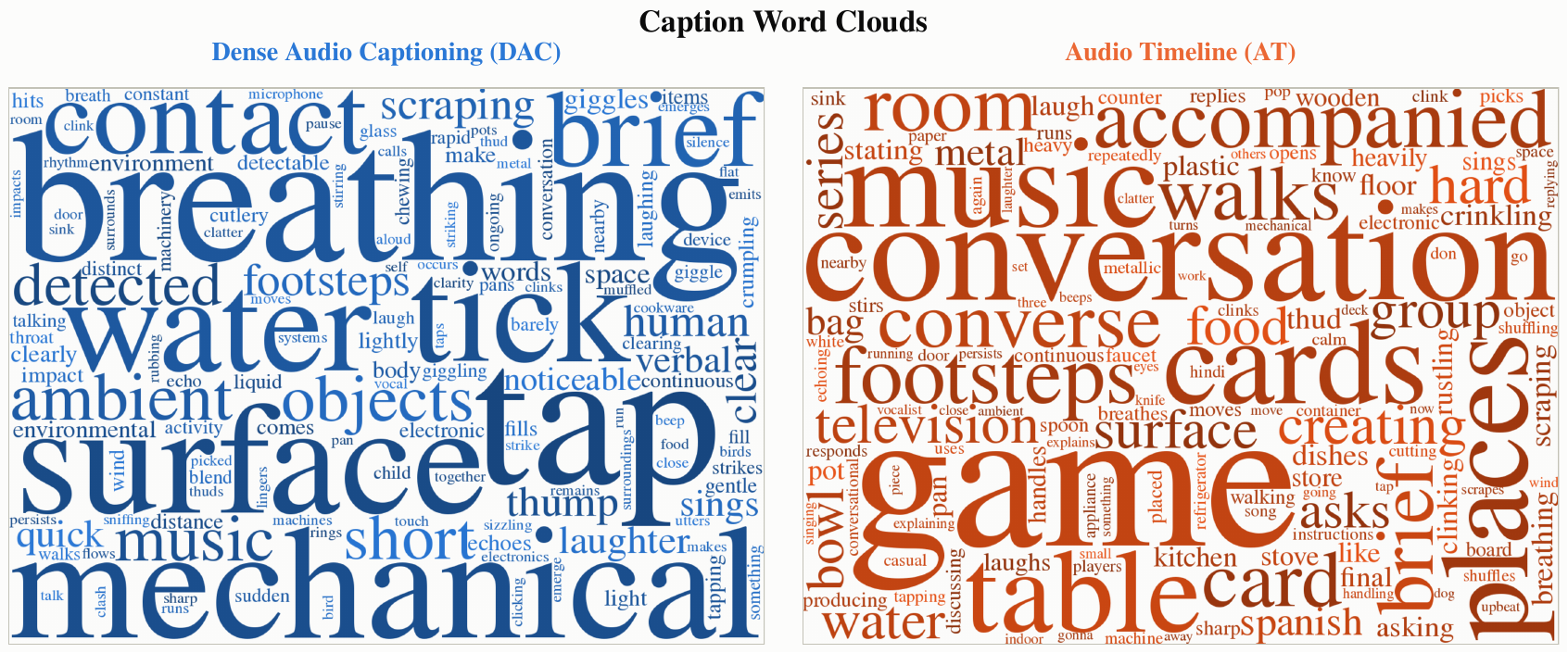}
        \caption{Word-clouds for DAC and AT tasks.}
        \label{subfig:egoscape_wordcloud}
    \end{subfigure}
    \hfill
    \begin{subfigure}[b]{0.4\textwidth}
        \includegraphics[width=\linewidth]{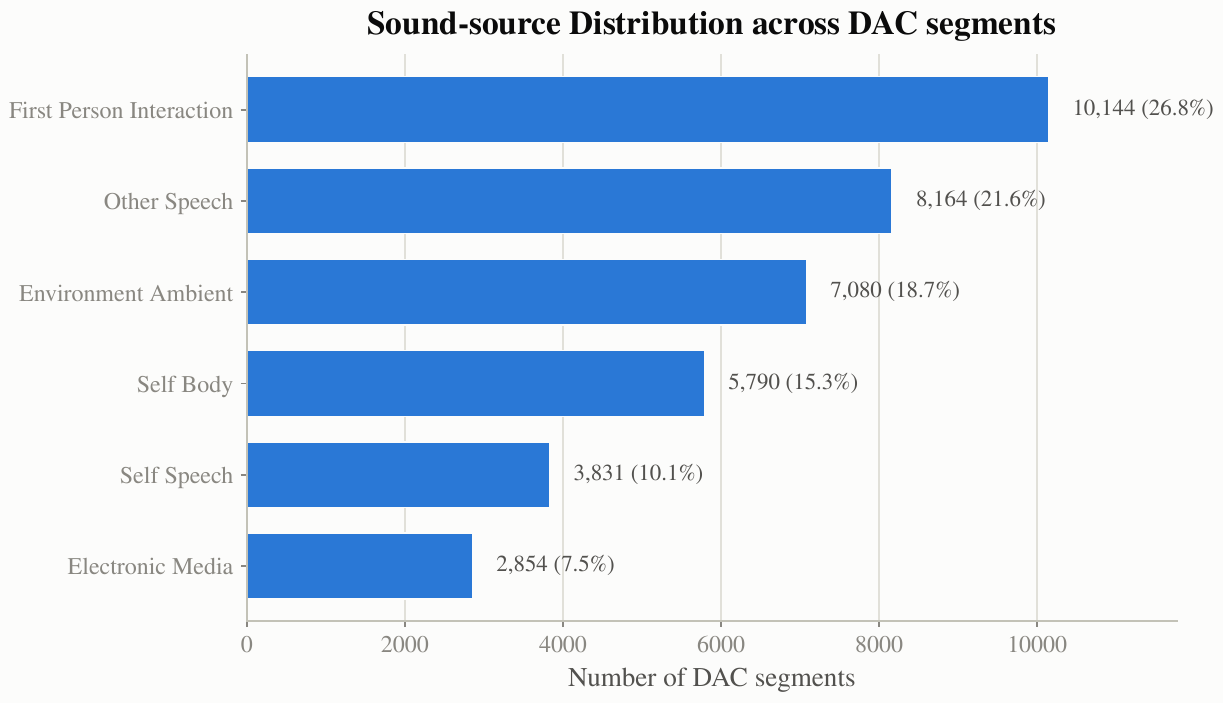}
        \caption{DAC sound source distribution.}
        \label{subfig:egoscape_dense_source_dist}
    \end{subfigure}

    \caption{Statistics for the EgoSCAPE benchmark tasks.}
    \label{fig:egoscape_benchmark_stats}
\end{figure}

\cref{fig:egoscape_benchmark_stats} shows the detailed stats for the EgoSCAPE benchmark. Our benchmark consists of 700 audio clips totaling 13.15 hours, with individual clips spanning durations of up to 5 minutes with a mean duration of around 67 seconds. The dense audio captions consist of a total of 37,803 segments with annotations about the detailed audio event present on it, with around 54 segments annotated per audio on average. Dense caption segments are about 2.7 seconds long on average and have a density of about 0.82 segments per second. On the other hand, the audio timeline consists of coarser time segments with a total of 4,587 segments and about 6.6 segments annotated per audio clip. These segments are 10 seconds long on average. \cref{subfig:egoscape_wordcloud} shows the word clouds for the dense captioning and audio timeline annotations. Finally, the distribution of the sound source annotations present in the dense captions is available in \cref{subfig:egoscape_dense_source_dist}.

\subsection{Human Verification Details}
\label{appsubsec:human_verification_details}

\begin{figure}
    \centering
    \includegraphics[width=\linewidth]{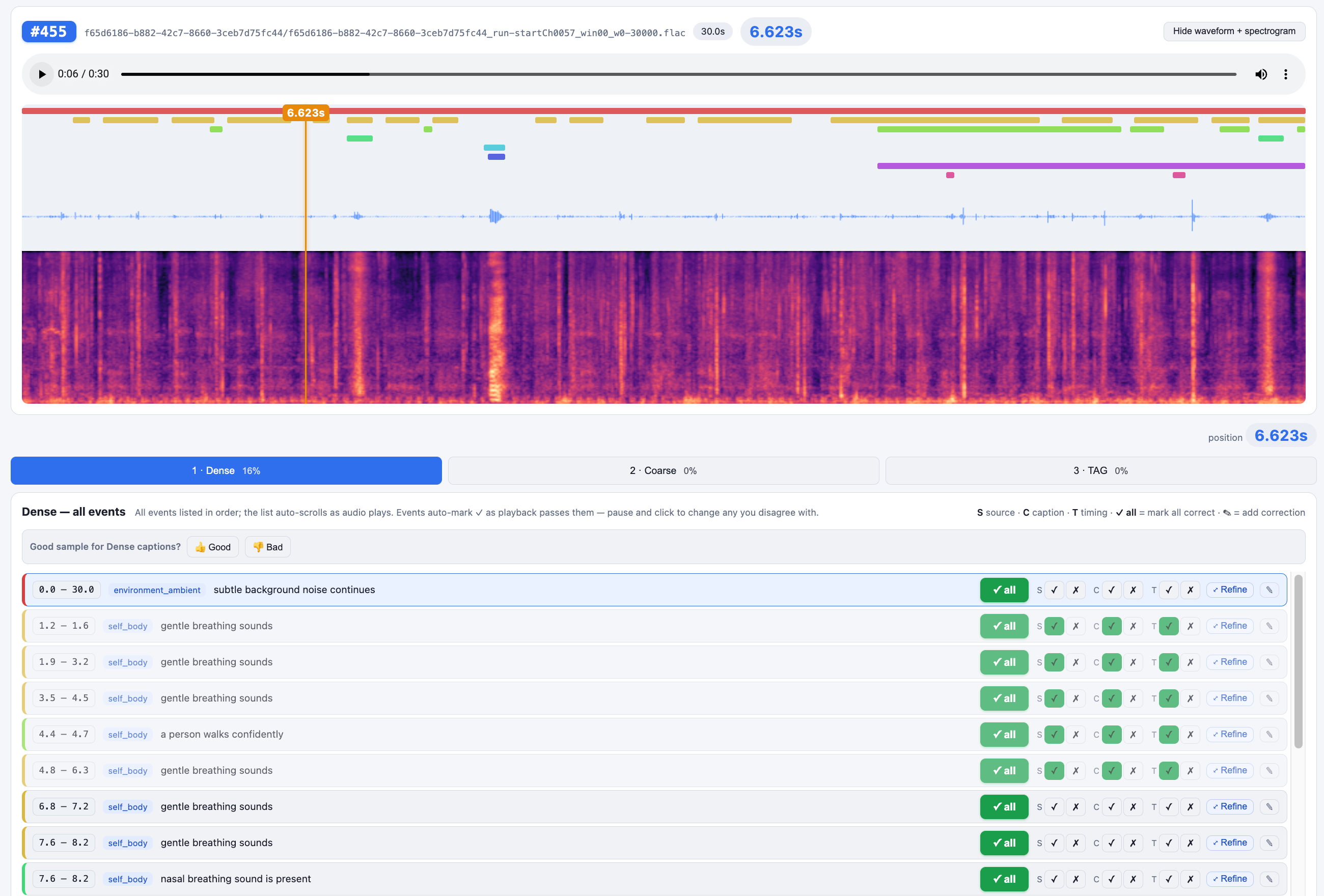}
    \caption{Dashboard used for performing human verification.}
    \label{fig:human_verification_dashboard}
\end{figure}

\begin{table}[]
\centering
\caption{Inter-annotator agreement for the human verification of EgoSCAPE benchmark. The agreement is computed over the fine-grained entities for each task: each time-stamped segment in dense captioning and audio timeline generation, and each query in temporal audio grounding.}
\label{tab:inter_annotator_agreement}
\resizebox{\columnwidth}{!}{%
\begin{tabular}{l|ccccc}
\hline\hline
\multicolumn{1}{c|}{\textbf{Task}} & \textbf{\# audios} & \textbf{\# entities} & \multicolumn{1}{l}{\textbf{\# annotators}} & \textbf{Raw Agree.} & \textbf{Fleiss' $\kappa$} \\ \hline
Dense Audio Captioning & 700 & 37803 & 3 & 90.7\% & 0.789 \\
Audio Timeline Generation & 700 & 4587 & 3 & 88.6\% & 0.764 \\
Temporal Audio Grounding & 700 & 2800 & 3 & 91.0\% & 0.797 \\ \hline\hline
\end{tabular}%
}
\end{table}

\cref{fig:human_verification_dashboard} shows the dashboard used for human verification. For each audio clip, annotators reviewed the annotations generated for the different tasks and marked whether the caption, the timing, and the associated sound sources of each audio segment were correct. If an annotator judged an annotation to be incorrect, they could also suggest a refined version. Annotators could play the audio files but had no access to the corresponding videos. To ensure accurate temporal annotations, the interface displayed both the audio waveform and its spectrogram.

\cref{tab:inter_annotator_agreement} reports the inter-annotator agreement for the human verification of the EgoSCAPE benchmark. Agreement is computed over the entities in each task, where an entity is a single time-stamped caption for dense captioning and timeline generation, and a single query for temporal grounding. An entity is considered incorrectly annotated by the automatic pipeline if an annotator marks its caption/query, timing, or sound source as wrong. We compute Fleiss' kappa ($\kappa$)~\citep{Fleiss1971kappa} over the annotators' judgments of whether each entity's annotations are correct.

\clearpage

\section{Experimental Details}
\label{appsec:experimental_details}

\subsection{Training details}
\label{appsubsec:training_details}

We adapt GRPO with the rewards and techniques described in \S\ref{sec:attest_method}. We train only the linear modules of the base model using LoRA~\citep{hu2022lora} with rank $r = 8$ and $\alpha = 32$. Prior to RL, we perform a cold-start SFT stage with a learning rate of $5 \times 10^{-5}$ on dense captioning, audio timeline, and temporal audio grounding data constructed from a separate set of Ego4D audio clips using the pipeline described in \S\ref{sec:egoscape_data}. Additionally, following TimeAudio~\citep{wang2026listening_timeaudio_ftar}, we convert AudioSet-SL~\citep{jort2016audioset,shawn2021audioset_sl} into dense captioning and temporal audio grounding formats and include it in the training mix. In total, we use 222k samples for DAC, 70k samples for AT, and 292k samples for temporal audio grounding. For RL fine-tuning, we use 5k samples each for the DAC and AT tasks. We set the GRPO group size to 8 and the KL regularization coefficient $\beta$ to 0.001. The LoRA adapter is fine-tuned with GRPO using a learning rate of $5 \times 10^{-6}$ and an effective batch size of 16. All experiments are conducted on 64 NVIDIA A100 (80GB) GPUs, and training is performed using the DeepSpeed\footnote{\url{https://www.deepspeed.ai/}} ZeRO-3 configuration.

To prevent the model from hacking $R_\text{AD}$ by predicting short segments, we cap $\rho$ at a maximum value of 5. Moreover, to encourage precise predictions, the format reward also penalizes outputs whose number of predicted segments differs from the number of ground-truth segments. In \S\ref{sec:attest_method}, a segment's $A^{\text{DAC}}$ term is $0$ if it is unmatched ($\sigma^{(g)}(j) = 0$), and its $A^{\text{CAG}}$ term is $0$ if it lies outside all silenced windows. Finally, $\alpha$ is set to 0.5 in \cref{eq:hungarian_matching_score}.

\subsection{Evaluation metrics}
\label{appsubsec:evaluation_metrics}

\subsubsection{Captioning metrics}
\label{appsubsubsec:captioning_metrics}


\paragraph{FENSE$@\tau$.} We report the FENSE scores \citep{zhou2022can_fense} at the thresholds of 0.3, 0.5 and 0.7 similar to \citep{shao2026listening_lat_audio} for the captioning tasks (DAC and AT). As defined in \S\ref{sec:attest_method}, the ground-truth has $N_g$ triplets $D = \{(s_i, e_i, c_i)\}_{i=1}^{N_g}$ where $(s_i, e_i)$ are the start/end timestamps of an acoustic event described by caption $c_i$. Similarly, the model prediction is represented as $\hat{D} = \{(\hat{s}_j, \hat{e}_j, \hat{c}_j)\}_{j=1}^{N_p}$. For each GT segment $i$, we identify the predicted segment $j^*$ that has the highest IoU above $\tau$ with the GT segment, i.e.
\begin{equation}
    FENSE_i@\tau = 
    \begin{cases}
        FENSE(c_i, \hat{c}_{j^*}) &\text{if } \text{tIoU}(d_i, \hat{d}_{j^*}) > \tau \\
        0 &\text{otherwise}
    \end{cases}
\end{equation}
where $j^* = \argmax_{j} \text{tIoU}(d_i, \hat{d}_{j})$, and $d_i$ \& $\hat{d}_j$ represent the triplets $(s_i, e_i, c_i)$ \& $(\hat{s}_j, \hat{e}_j, \hat{c}_j)$ respectively. The final metric is the average over all the GT triplets, i.e.
\begin{equation}
    FENSE@\tau = \frac{1}{N_g} \sum_{i=1}^{N_g} FENSE_i@\tau
\end{equation}

\paragraph{SED F1 metrics.} For the DAC task we additionally report two sound-event-detection style F1 scores adapted from \citet{mesaros2016metrics_sed}: a frame-based \textbf{Seg-F1} and an event-based \textbf{Evt-F1} with an onset collar of $0.5$\,s. Since our captions are free-form rather than drawn from a fixed class vocabulary, we induce per-clip event classes by clustering the captions of $D$ with single-linkage connected components over the SBERT (paraphrase-TinyBERT-L6-v2) \citep{reimers2019sbert} cosine-similarity graph thresholded at $\tau_{\text{sim}}=0.65$. Let $\kappa(\cdot)$ denote the resulting class assignment, so that two captions are of the same class iff they land in the same cluster. For both metrics we report the micro-averaged F1 obtained by pooling the counts across all clips of the test set.

\paragraph{Hallucination Rate.} Finally we report an additional hallucination rate similar to TAC \citep{kumar2026tac} for the DAC task. We report the hallucination rate (Hal\%) as the percentage of predicted triplets or segments that are not matched to any of the GT segments at tIoU of 0.5 and have a \emph{confidence} of less than 0.15. \emph{Confidence} is defined as the maximum audio-text FLAM \citep{flam2025} similarity score across the time-region of a predicted segment/triplet.

\subsubsection{Temporal Audio Grounding Metrics}

\paragraph{Recall$@\tau$.} We compute the tIoU between the predicted time-ranges and the ground-truth time ranges and report the recall at the thresholds of 0.5 and 0.7 tIoU. 

\paragraph{mIoU.} We report the mean tIoU scores across all the samples of the dataset as mIoU.

\subsection{Baseline implementation details}
\label{appsec:baseline_details}

\subsubsection{LLM based models}

All models are asked to emit
timestamps in seconds at a 100\,ms (one-decimal) precision. For captioning tasks the instruction requests a flat list of \texttt{start - end, description} segments and for temporal audio grounding (TAG) it requests only the \texttt{start - end} ranges where the query occurs. The EgoSCAPE dense task additionally requests a source tag drawn from a taxonomy (\cref{subfig:egoscape_dense_source_dist}) prepended to each segment. Below we describe only what is \emph{specific} to each baseline.

\paragraph{Proprietary models (Gemini).}
The only model-specific
post-processing is timestamp normalization: Gemini tends to emit clock-style (\texttt{m:ss}) timestamps, which we convert to decimal seconds before scoring so that the shared parser within our evaluation pipeline reads them correctly.

\paragraph{Open audio and omni LLMs.}
These models split into two groups by how they handle long audio.
\emph{Native long-audio} models are run with a single forward pass and rely on their own internal sliding-window audio encoders: AudioFlamingo~2 (up to 5\,min), AudioFlamingo~3 (up to $\sim$600\,s), and AF-Next ($\sim$1800\,s). \emph{Chunk-and-merge} models are those whose audio front-end caps at 30\,s (Qwen2-Audio and Kimi-Audio). For these we split the clip into non-overlapping 30\,s windows, run inference on each window independently, offset every predicted timestamp by its window start, and merge the per-window predictions into a single clip-level prediction (unioning overlapping ranges for TAG and concatenating segments for DAC) before scoring, so the unchanged scorer sees one prediction per clip.


\paragraph{Captioning models.}
The captioner baselines (AF-Next Captioner and Qwen3-Omni Captioner) are run in two stages
because they are trained to produce rich free-form captions rather than a rigid task format. First, the audio model is prompted to produce a timestamped free-form caption of the clip. Second, a text-only LLM (Qwen3-30B-Instruct) reformats that caption into the exact DAC/timeline/TAG format, using the corresponding task instruction as the formatting specification. Because Qwen3 Omni Captioner captions best on short inputs, it is also chunked into 30\,s windows and the reformatting stage is applied per window with a deterministic offset added to each window's timestamps before concatenation, so all timeline arithmetic is exact.

\paragraph{Specialized temporal models.}
For longer clips processed through TimeAudio, we apply the same windowed mechanism as above with 10\,s non-overlapping windows; each window's timestamps are shifted by its start and stitched into one prediction. TimeAudio results obtained this way are marked with $*$ in the main tables. LAT-Audio is run with its tool-augmented reasoning loop, in which the model iteratively issues audio-crop calls to zoom into sub-segments before answering, and thus handles long audio natively without fixed chunking. SpotSound-A is used off-the-shelf as it is an AudioFlamingo-3-based grounding model with timestamp-interleaved audio tokens that reads the query and emits a single grounded interval. AM-DETR is the baseline from \citet{munakata2025language_clotho_moment_lamr} and UVCOM is the baseline from \citet{munakata2026castella}.

\subsubsection{RL-based baselines}

We also adapt TimePro-RL \citep{shi2026towards_timepro_rl} for our setup by using identical training data and models. We include temporal markers every 0.04 seconds similar to their approach causing the context usage to increase by 7500 tokens for a 300 seconds audio. Additionally, as suggested by TimePro-RL, we include the main reward to be the task metric and auxiliary reward as $R_\text{DAC}$ as we have defined.

\subsection{Figure 3 details}
\label{appsubsec:figure3_details}

For \cref{fig:hallucination_sft}, we plot the hallucination rate for the SFT trained variants of Qwen3 Omni and Qwen2.5 Omni, over different durations of audios from EgoSCAPE benchmark. We can see in the figure that the hallucination rate clearly rises with the duration of the audio for both the SFT trained models.

\clearpage

\section{Extended Results}
\label{appsec:extended_results}

\subsection{Audio captioning}
\label{appsubsec:audio_captioning_results}

\begin{table}[h]
\centering
\caption{Dense audio captioning results (expanded from \cref{tab:dense_cap_results}) on EgoSCAPE and AudioSet-SL datasets comparing ATTEST to other audio and omni LLM baselines. EvtF1@0.5 and SegF1 are micro. Hal\% and SrcF1@0.5 are macro.}
\label{apptab:dense_cap_results_extended}
\resizebox{\columnwidth}{!}{%
\begin{tabular}{l|ccccccc|cccccc}
\hline \hline
\multicolumn{1}{c|}{\multirow{2}{*}{\textbf{Model}}} & \multicolumn{7}{c|}{\textbf{EgoSCAPE Dense Audio Captioning}} & \multicolumn{6}{c}{\textbf{AudioSet-SL}} \\ \cline{2-14} 
\multicolumn{1}{c|}{} & \textbf{F@0.3} & \textbf{F@0.5} & \textbf{F@0.7} & \textbf{EvtF1} & \textbf{SegF1} & \textbf{Hal\%$\downarrow$} & \textbf{SrcF1} & \textbf{F@0.3} & \textbf{F@0.5} & \textbf{F@0.7} & \textbf{EvtF1} & \textbf{SegF1} & \textbf{Hal\%$\downarrow$} \\ \hline
\multicolumn{14}{c}{\emph{Proprietary Models}} \\ \hline
Gemini 3.1 Pro & 0.235 & 0.213 & 0.201 & 0.084 & 0.276 & 31.1 & 0.371 & 0.207 & 0.192 & 0.181 & 0.079 & 0.126 & 22.7 \\
Gemini 3.5 Flash & 0.226 & 0.208 & 0.198 & 0.066 & 0.180 & 38.5 & 0.327 & 0.208 & 0.192 & 0.176 & 0.043 & 0.074 & 23.7 \\
Gemini 3.6 Flash & 0.242 & 0.215 & 0.200 & 0.091 & 0.197 & 38.9 & 0.351 & 0.217 & 0.200 & 0.172 & 0.037 & 0.066 & 30.3 \\
Gemini 3.1 Flash Lite & 0.228 & 0.211 & 0.199 & 0.064 & 0.149 & 34.7 & 0.400 & 0.202 & 0.185 & 0.171 & 0.032 & 0.064 & 25.1 \\
GPT RealTime 2.1 & 0.207 & 0.199 & 0.195 & 0.026 & 0.113 & 42.2 & 0.387 & 0.193 & 0.176 & 0.165 & 0.026 & 0.040 & 54.5 \\
GPT RealTime 2.0 & 0.209 & 0.200 & 0.196 & 0.025 & 0.114 & 45.6 & 0.387 & 0.208 & 0.186 & 0.170 & 0.038 & 0.045 & 52.9 \\
GPT 4o Audio & 0.226 & 0.208 & 0.198 & 0.059 & 0.186 & 48.4 & 0.359 & 0.208 & 0.184 & 0.167 & 0.051 & 0.058 & 50.9 \\\hline
\multicolumn{14}{c}{\emph{Captioning Models}} \\ \hline
AF-Next Captioner & 0.244 & 0.218 & 0.201 & 0.099 & 0.147 & 52.7 & 0.325 & 0.239 & 0.207 & 0.184 & 0.058 & 0.087 & 44.6 \\
Qwen3 Omni Captioner & 0.270 & 0.232 & 0.206 & 0.257 & 0.335 & 64.4 & 0.369 & 0.257 & 0.220 & 0.190 & 0.048 & 0.057 & 63.6 \\ \hline 
\multicolumn{14}{c}{\emph{Audio and Omni LLMs}} \\ \hline
Qwen2 Audio* & 0.207 & 0.199 & 0.196 & 0.048 & 0.127 & 51.0 & 0.232 & 0.207 & 0.191 & 0.182 & 0.054 & 0.076 & 43.1 \\
Kimi-Audio* & 0.188 & 0.183 & 0.177 & 0.042 & 0.034 & 42.9 & 0.207 & 0.224 & 0.198 & 0.175 & 0.061 & 0.107 & 33.9 \\
TimeAudio* & 0.340 & 0.310 & 0.271 & 0.256 & 0.344 & 35.2 & - & 0.426 & 0.389 & 0.352 & 0.353 & 0.372 & 36.3 \\
AudioFlamingo 2 & 0.193 & 0.193 & 0.192 & 0.004 & 0.001 & 59.7 & - & 0.192 & 0.176 & 0.167 & 0.079 & 0.084 & 55.5 \\
AF-Next & 0.181 & 0.178 & 0.176 & 0.022 & 0.014 & 45.9 & 0.030 & 0.226 & 0.202 & 0.184 & 0.089 & 0.114 & 48.5 \\ \hline
Qwen2.5 Omni & 0.178 & 0.177 & 0.176 & 0.021 & 0.035 & 48.2 & 0.165 & 0.220 & 0.202 & 0.192 & 0.124 & 0.190 & 20.5 \\
+ SFT & 0.412 & 0.367 & 0.313 & 0.320 & 0.303 & 47.2 & 0.564 & 0.448 & 0.403 & 0.362 & 0.336 & 0.337 & 25.8 \\
+ TimePro-RL & 0.445 & 0.410 & 0.358 & 0.349 & 0.348 & 31.3 & 0.642 & 0.475 & 0.430 & 0.385 & 0.347 & 0.370 & 20.1 \\
\rowcolor{metalight}
\textbf{+ ATTEST (Ours)} & \textbf{0.461} & \textbf{0.427} & \textbf{0.372} & \textbf{0.371} & \textbf{0.355} & \textbf{30.5} & \textbf{0.656} & \textbf{0.489} & \textbf{0.442} & \textbf{0.399} & \textbf{0.354} & 0.389 & \textbf{18.2} \\ \hline
Qwen3 Omni & 0.178 & 0.176 & 0.176 & 0.015 & 0.038 & 45.4 & 0.231 & 0.265 & 0.231 & 0.208 & 0.138 & 0.175 & 29.2 \\
+ SFT & 0.419 & 0.366 & 0.318 & 0.301 & 0.325 & 47.8 & 0.547 & 0.438 & 0.396 & 0.368 & 0.371 & 0.340 & 34.5 \\
+ TimePro-RL & 0.451 & 0.415 & 0.359 & 0.382 & 0.337 & 32.5 & 0.658 & 0.484 & 0.438 & 0.389 & 0.399 & 0.411 & 28.0 \\
\rowcolor{metalight}
\textbf{+ ATTEST (Ours)} & \textbf{0.463} & \textbf{0.428} & \textbf{0.373} & \textbf{0.400} & \textbf{0.353} & \textbf{30.7} & \textbf{0.665} & \textbf{0.498} & \textbf{0.450} & \textbf{0.404} & \textbf{0.410} & \textbf{0.425} & \textbf{26.9} \\ \hline
AudioFlamingo 3 & 0.196 & 0.194 & 0.193 & 0.042 & 0.010 & 40.3 & - & 0.204 & 0.189 & 0.177 & 0.119 & 0.131 & 32.4 \\
+ SFT & 0.376 & 0.344 & 0.298 & 0.258 & 0.248 & 44.4 & 0.587 & 0.453 & 0.406 & 0.362 & 0.345 & 0.396 & 36.3 \\
+ TimePro-RL & 0.421 & 0.388 & 0.328 & 0.335 & 0.329 & 34.0 & 0.620 & 0.489 & 0.440 & 0.384 & 0.369 & 0.385 & 29.2 \\
\rowcolor{metalight}
\textbf{+ ATTEST (Ours)} & \textbf{0.437} & \textbf{0.400} & \textbf{0.342} & \textbf{0.346} & \textbf{0.345} & \textbf{31.8 }& \textbf{0.641} & \textbf{0.503} & \textbf{0.451} & \textbf{0.398} & \textbf{0.380} & \textbf{0.400} & \textbf{25.3 }\\ 
\hline \hline
\end{tabular}%
}
\end{table}

\noindent \textbf{Dense audio captioning.} \cref{apptab:dense_cap_results_extended} is an expanded version of the DAC results present in \S\ref{subsec:results_captioning} including ATTEST framework results on Qwen3 Omni. Additionally, we report results on more proprietary baseline models here including Gemini 3.5 Flash, Gemini 3.1 Flash Lite, GPT RealTime 2.1 and GPT RealTime 2.0. These additional results further strengthen the effectiveness of the proposed approach for dense captioning.

\begin{table}[]
\centering
\caption{Audio timeline generation results on the EgoSCAPE benchmark.}
\label{tab:egoscape_timeline_results}
\resizebox{0.5\columnwidth}{!}{%
\begin{tabular}{l|ccc}
\hline \hline
\multicolumn{1}{c|}{\textbf{Model}} & \textbf{F@0.3} & \textbf{F@0.5} & \textbf{F@0.7} \\ \hline
\multicolumn{4}{c}{\emph{Proprietary Models}} \\ \hline
Gemini 3.1 Pro & 0.234 & 0.197 & 0.172 \\
Gemini 3.5 Flash & 0.213 & 0.185 & 0.166 \\
Gemini 3.6 Flash & 0.218 & 0.183 & 0.164 \\
Gemini 3.1 Flash Lite & 0.245 & 0.205 & 0.174 \\ 
GPT RealTime 2.1 & 0.209 & 0.184 & 0.167 \\
GPT RealTime 2.0 & 0.217 & 0.190 & 0.166 \\
GPT 4o Audio & 0.239 & 0.199 & 0.166 \\\hline
\multicolumn{4}{c}{\emph{Captioning Models}} \\ \hline
AF-Next Captioner & 0.241 & 0.199 & 0.169 \\
Qwen3 Omni Captioner & 0.341 & 0.258 & 0.185 \\ \hline
\multicolumn{4}{c}{\emph{Audio and Omni LLMs}} \\ \hline
Qwen2 Audio & 0.153 & 0.150 & 0.149 \\
Kimi-Audio & 0.186 & 0.170 & 0.161 \\
AudioFlamingo 2 & 0.164 & 0.156 & 0.154 \\
AF-Next & 0.183 & 0.176 & 0.160 \\ 
LAT-Audio & 0.220 & 0.184 & 0.168 \\ \hline
Qwen2.5 Omni & 0.211 & 0.189 & 0.170 \\
{+ SFT} & 0.466 & 0.401 & 0.316 \\
{+ TimePro-RL} & 0.489 & 0.427 & 0.329 \\
\rowcolor{metalight}
\textbf{+ ATTEST (Ours)} & \textbf{0.503} & \textbf{0.436} & \textbf{0.340} \\ \hline
Qwen3 Omni & 0.304 & 0.255 & 0.208 \\
{+ SFT} & 0.469 & 0.403 & 0.313 \\
{+ TimePro-RL} & 0.495 & 0.429 & 0.332 \\
\rowcolor{metalight}
\textbf{+ ATTEST (Ours)} & \textbf{0.509} & \textbf{0.442} & \textbf{0.345} \\ \hline
AudioFlamingo 3 & 0.187 & 0.168 & 0.159 \\
{+ SFT} & 0.458 & 0.396 & 0.312 \\
{+ TimePro-RL} & 0.489 & 0.425 & 0.329 \\
\rowcolor{metalight}
\textbf{+ ATTEST (Ours)} & \textbf{0.505} & \textbf{0.438} & \textbf{0.341} \\ \hline \hline
\end{tabular}%
}
\end{table}

\noindent \textbf{Audio timeline generation.} 
\cref{tab:egoscape_timeline_results} similarly reports the performance comparison between the baselines and proposed approach for the audio timeline generation task of EgoSCAPE benchmark. The ATTEST framework significantly improves the performance of the base models for this task as well outperforming all the baselines. While the performance of our base models and finetuned models are better on timeline generation compared to dense captioning, the performance of some baselines such as Gemini 3.1 Pro and AudioFlamingo models is similar.

\subsection{Temporal audio grounding}
\label{appsubsec:tag_results_extended}

\begin{table}[]
\centering
\caption{Zero-shot temporal audio grounding results on Clotho-Moment, UnAV-100 and CASTELLA datasets comparing ATTEST to other audio and omni LLM baselines. }
\label{apptab:tag_results_zeroshot_expanded}
\resizebox{\columnwidth}{!}{%
\begin{tabular}{l|ccc|ccc|ccc}
\hline \hline
\multicolumn{1}{c|}{\multirow{2}{*}{\textbf{Model}}} & \multicolumn{3}{c|}{\textbf{Clotho-Moment}} & \multicolumn{3}{c|}{\textbf{UnAV-100}} & \multicolumn{3}{c}{\textbf{CASTELLA}} \\ \cline{2-10} 
\multicolumn{1}{c|}{} & \textbf{R@0.5} & \textbf{R@0.7} & \textbf{mIoU} & \textbf{R@0.5} & \textbf{R@0.7} & \textbf{mIoU} & \textbf{R@0.5} & \textbf{R@0.7} & \textbf{mIoU} \\ \hline
\multicolumn{10}{c}{\emph{Proprietary Models}} \\ \hline
Gemini 3.1 Pro & 0.385 & 0.296 & 0.344 & 0.340 & 0.240 & 0.294 & 0.080 & 0.041 & 0.081 \\
Gemini 3.5 Flash & 0.189 & 0.079 & 0.223 & 0.430 & 0.220 & 0.373 & 0.082 & 0.039 & 0.098 \\
Gemini 3.6 Flash & 0.581 & 0.431 & 0.536 & 0.570 & 0.410 & 0.524 & 0.233 & 0.124 & 0.223 \\
Gemini 3.1 Flash Lite & 0.235 & 0.134 & 0.302 & 0.540 & 0.340 & 0.487 & 0.042 & 0.021 & 0.050 \\ 
GPT RealTime 2.1 & 0.021 & 0.007 & 0.056 & 0.090 & 0.030 & 0.135 & 0.029 & 0.013 & 0.049 \\
GPT RealTime 2.0 & 0.042 & 0.011 & 0.085 & 0.100 & 0.070 & 0.166 & 0.036 & 0.012 & 0.053 \\ 
GPT 4o Audio & 0.106 & 0.031 & 0.185 & 0.200 & 0.060 & 0.216 & 0.042 & 0.018 & 0.062 \\ \hline
\multicolumn{10}{c}{\emph{Finetuned Temporal Grounding Models}} \\ \hline
AM-DETR & 0.882 & 0.827 & 0.809 & 0.580 & 0.410 & 0.428 & - & - & - \\ 
SpotSound-A & 0.908 & 0.853 & 0.868 & 0.780 & 0.620 & 0.725 & 0.180 & 0.115 & 0.181 \\
UVCOM & 0.885 & 0.832 & 0.820 & 0.590 & 0.360 & 0.518 & \underline{0.302} & \underline{0.191} & \underline{0.317} \\ \hline
\multicolumn{10}{c}{\emph{Audio and Omni LLMs}} \\ \hline
Qwen2 Audio* & 0.043 & 0.005 & 0.131 & 0.080 & 0.040 & 0.181 & 0.053 & 0.020 & 0.097 \\
Kimi-Audio* & 0.088 & 0.025 & 0.164 & 0.180 & 0.030 & 0.265 & 0.080 & 0.022 & 0.146 \\
AudioFlamingo 2 & 0.009 & 0.003 & 0.023 & 0.040 & 0.010 & 0.056 & 0.011 & 0.005 & 0.016 \\
AF-Next & 0.004 & 0.002 & 0.019 & 0.030 & 0.030 & 0.069 & 0.008 & 0.002 & 0.013 \\ \hline
Qwen2.5 Omni & 0.409 & 0.141 & 0.426 & 0.470 & 0.280 & 0.458 & 0.092 & 0.048 & 0.116 \\
+ SFT & 0.875 & 0.821 & 0.830 & 0.790 & 0.600 & 0.693 & 0.224 & 0.123 & 0.269 \\
+ TimePro-RL & 0.895 & 0.844 & 0.859 & 0.800 & 0.620 & 0.711 & 0.252 & 0.143 & 0.295 \\
\rowcolor{metalight}
\textbf{+ ATTEST (Ours)} & \textbf{0.909} & \textbf{0.859} & \textbf{0.876} & \textbf{0.820} & \textbf{0.650} & \textbf{0.729} & \textbf{0.297} & \textbf{0.186} & \textbf{0.312} \\ \hline
Qwen3 Omni & 0.495 & 0.424 & 0.528 & 0.760 & 0.590 & 0.681 & 0.190 & 0.109 & 0.187 \\
+ SFT & 0.872 & 0.828 & 0.846 & 0.790 & 0.610 & 0.717 & 0.207 & 0.136 & 0.229 \\
+ TimePro-RL & 0.890 & 0.834 & 0.860 & 0.810 & 0.620 & 0.731 & 0.219 & 0.139 & 0.245 \\
\rowcolor{metalight}
\textbf{+ ATTEST (Ours)} & \textbf{0.904} & \textbf{0.841} & \textbf{0.871} & \textbf{0.830} & \textbf{0.640} & \textbf{0.745} & \textbf{0.264} & \textbf{0.155} & \textbf{0.298} \\ \hline
AudioFlamingo 3 & 0.142 & 0.063 & 0.163 & 0.120 & 0.030 & 0.147 & 0.008 & 0.004 & 0.013 \\
+ SFT & 0.850 & 0.754 & 0.801 & 0.780 & 0.620 & 0.708 & 0.147 & 0.095 & 0.170 \\
+ TimePro-RL & 0.892 & 0.799 & 0.836 & 0.810 & 0.650 & 0.744 & 0.168 & 0.110 & 0.189 \\
\rowcolor{metalight}
\textbf{+ ATTEST (Ours)} & \textbf{0.903} & \textbf{0.819} & \textbf{0.854} & \textbf{0.850} & \textbf{0.660} & \textbf{0.753} & \textbf{0.202} & \textbf{0.123} & \textbf{0.266} \\ 
\hline \hline
\end{tabular}%
}
\end{table}

\begin{table}[t]
\centering
\caption{Temporal audio grounding results on EgoSCAPE benchmark across coarse and fine-grained audio-related queries.}
\label{tab:tag_results_egoscape}
\resizebox{0.7\columnwidth}{!}{%
\begin{tabular}{l|ccc|ccc}
\hline \hline
\multicolumn{1}{c|}{\multirow{2}{*}{\textbf{Model}}} & \multicolumn{3}{c|}{\textbf{Coarse}} & \multicolumn{3}{c}{\textbf{Fine}} \\ \cline{2-7} 
\multicolumn{1}{c|}{} & \textbf{R@0.5} & \textbf{R@0.7} & \textbf{mIoU} & \textbf{R@0.5} & \textbf{R@0.7} & \textbf{mIoU} \\ \hline
\multicolumn{7}{c}{\emph{Proprietary Models}} \\ \hline
Gemini 3.1 Pro & 0.173 & 0.110 & 0.191 & 0.060 & 0.042 & 0.077 \\
Gemini 3.5 Flash & 0.159 & 0.077 & 0.192 & 0.048 & 0.028 & 0.065 \\
Gemini 3.6 Flash & 0.197 & 0.106 & 0.236 & 0.077 & 0.046 & 0.094 \\
Gemini 3.1 Flash Lite & 0.184 & 0.117 & 0.217 & 0.074 & 0.048 & 0.090 \\
GPT RealTime 2.1 & 0.072 & 0.033 & 0.112 & 0.023 & 0.007 & 0.033 \\
GPT RealTime 2.0 & 0.103 & 0.045 & 0.131 & 0.032 & 0.017 & 0.045 \\
GPT 4o Audio & 0.077 & 0.032 & 0.135 & 0.036 & 0.011 & 0.056 \\\hline
\multicolumn{7}{c}{\emph{Audio and Omni LLMs}} \\ \hline
Qwen2 Audio* & 0.092 & 0.045 & 0.126 & 0.027 & 0.011 & 0.039 \\
Kimi-Audio* & 0.123 & 0.049 & 0.168 & 0.072 & 0.019 & 0.098 \\
SpotSound-A & 0.279 & 0.155 & 0.283 & 0.172 & 0.106 & 0.183 \\
AudioFlamingo 2 & 0.053 & 0.011 & 0.069 & 0.025 & 0.010 & 0.041 \\
AF-Next & 0.028 & 0.015 & 0.044 & 0.025 & 0.013 & 0.034 \\ \hline
Qwen2.5 Omni & 0.202 & 0.105 & 0.215 & 0.069 & 0.043 & 0.084 \\
+ SFT & 0.444 & 0.307 & 0.436 & 0.506 & 0.366 & 0.467\\
+ TimePro-RL & 0.492 & 0.339 & 0.489 & 0.510 & 0.375 & 0.474 \\
\rowcolor{metalight}
\textbf{+ ATTEST (Ours)} & \textbf{0.518} & \textbf{0.358} & \textbf{0.505} & \textbf{0.529} & \textbf{0.393} & \textbf{0.481} \\ \hline
Qwen3 Omni & 0.243 & 0.169 & 0.280 & 0.119 & 0.070 & 0.132 \\
+ SFT & 0.443 & 0.302 & 0.431 & 0.458 & 0.346 & 0.426 \\
+ TimePro-RL & 0.521 & 0.326 & 0.505 & 0.498 & 0.361 & 0.456 \\
\rowcolor{metalight}
\textbf{+ ATTEST (Ours)} & \textbf{0.539} & \textbf{0.343} & \textbf{0.521} & \textbf{0.514} & \textbf{0.378} & \textbf{0.479} \\ \hline
AudioFlamingo 3 & 0.151 & 0.071 & 0.167 & 0.073 & 0.033 & 0.091 \\
+ SFT & 0.393 & 0.279 & 0.403 & 0.398 & 0.285 & 0.367 \\
+ TimePro-RL & 0.538 & 0.355 & 0.539 & 0.460 & 0.328 & 0.428 \\
\rowcolor{metalight}
\textbf{+ ATTEST (Ours)} & \textbf{0.574} & \textbf{0.376} & \textbf{0.556} & \textbf{0.475} & \textbf{0.334} & \textbf{0.447} \\ \hline \hline
\end{tabular}%
}
\end{table}

\noindent\textbf{Zero-shot results.} \cref{apptab:tag_results_zeroshot_expanded} is an expanded version of the TAG results present in \S\ref{subsec:results_grounding} including the ATTEST framework results on Qwen3 Omni. Similar to \cref{apptab:dense_cap_results_extended}, we include the results from additional proprietary models on different benchmarks.

\noindent\textbf{EgoSCAPE.} For the proposed EgoSCAPE benchmark, we report the temporal grounding results in \cref{tab:tag_results_egoscape} over coarse and fine-grained audio queries (\S\ref{subsec:automatic_benchmark_creation}). While the baselines are particularly poor at grounding fine-grained audio events, base models tuned with the proposed approach are equally good at handling coarse and fine-grained queries.

\begin{table}[]
\centering
\caption{Results of the proposed approach on LongAudioBench and MMAR.}
\label{apptab:general_audio_benchmarks_expanded}
\resizebox{0.6\columnwidth}{!}{%
\begin{tabular}{l|cc|cc}
\hline\hline
\multicolumn{1}{c|}{\multirow{2}{*}{\textbf{Model}}} & \multicolumn{2}{c|}{\textbf{MMAR}} & \multicolumn{2}{c}{\textbf{LongAudioBench}} \\ \cline{2-5} 
\multicolumn{1}{c|}{} & \textbf{All} & \textbf{Temporal} & \textbf{All} & \textbf{Temporal} \\ \hline
Qwen2.5 Omni & 63.2 & 56.0 & 61.5 & 60.1 \\
+ TimePro-RL & 61.0 & 55.4 & 58.3 & 59.5 \\
\rowcolor{metalight}
+ \textbf{ATTEST (Ours)} & \textbf{64.4} & \textbf{58.2} & \textbf{62.8} & \textbf{62.2} \\
+ ATTEST w/ markers & 62.2 & 56.9 & 60.5 & 61.2 \\ \hline
Qwen3 Omni & 73.4 & 67.1 & 65.9 & 64.8 \\
+ TimePro-RL & 71.0 & 65.8 & 62.3 & 62.7 \\
\rowcolor{metalight}
+ \textbf{ATTEST (Ours)} & \textbf{74.1} & \textbf{69.2} & \textbf{66.5} & \textbf{65.9} \\
+ ATTEST w/ markers & 72.8 & 66.5 & 64.1 & 64.0 \\ \hline
AudioFlamingo 3 & 60.4 & 53.2 & 57.7 & 53.9 \\
+ TimePro-RL & 58.3 & 51.0 & 55.2 & 52.5 \\
\rowcolor{metalight}
+ \textbf{ATTEST (Ours)} & \textbf{62.1} & \textbf{57.9} & \textbf{59.6} & \textbf{57.2} \\
+ ATTEST w/ markers & 59.2 & 52.8 & 56.9 & 53.5 \\\hline\hline
\end{tabular}%
}
\end{table}

\subsection{General audio benchmarks}
\label{appsubsec:general_audio_benchmark_results}

\cref{apptab:general_audio_benchmarks_expanded} expands the results reported on general benchmarks in \S\ref{subsec:results_general_benchmarks}. We include results of different post-training techniques including ATTEST on Qwen3 Omni. 


\subsection{Hyper-parameter and reward design}
\label{subsubsec:hyperparam_reward_design}

This subsection extends the ablations and hyper-parameter analysis from \S\ref{sec:ablations}.

\textbf{Reward weight tuning.} \cref{tab:reward_weight_sweep} reports the effect of varying the advantage weights $\eta_1$, $\eta_2$, and $\eta_3$ one at a time around our default setting $(\eta_1, \eta_2, \eta_3) = (1, 0.2, 1)$. Performance on EgoSCAPE dense captioning and Clotho-Moment is stable across the sweep: F@0.5 stays within 0.410--0.427 and Clotho-Moment R@0.5 within 0.891--0.909. CASTELLA is more sensitive, with R@0.5 ranging from 0.250 to 0.297. Every configuration in the sweep still outperforms vanilla GRPO (\cref{tab:ablation_reward}), so ATTEST does not depend on precise tuning of these weights. The trends also match the role of each reward. Lowering $\eta_1$ ($R_\text{CAG}$) increases hallucinations the most (35.5\% at $\eta_1 = 0.25$). Raising $\eta_2$ ($R_\text{AD}$) to 0.5 gives the lowest hallucination rate (29.5\%) but reduces F@0.5, indicating a trade-off between hallucination rate and caption quality. Reducing $\eta_3$ ($R_\text{DAC}$) lowers both captioning and grounding performance.

\textbf{Counterfactual Audio Grounding Reward.} For $R_\text{CAG}$, we report different variations in the corruption operator $M(\cdot)$ as well as the fraction of the audio clip where this corruption is applied to in \cref{tab:r_cag_modification_sweep}. In addition to silence, we tried other corruption operators such as pink noise (RMS-matched to the original content), time-shuffle, audio swap as well as using a low-pass filter. We show that the silence operator works the best in reducing model hallucinations and improving performance, as it is the only signal that converts the audio to something that truly does not contain any acoustic events. Moreover, changing the corruption fraction to high values of 0.5 reduces the model performance significantly due to over-penalization.

\textbf{Attention Density Reward.} We show the effect of using different layers to drive $R_\text{AD}$ in \cref{tab:r_ad_attention_layer}. We report the performance on using the first, middle, and last layers of the base Qwen2.5 Omni model. Moreover, we also tried a combination of the first, middle and last layers as well as using every 3 layers for $R_\text{AD}$. Finally, we also experimented by using the layer which has the peak difference between the attention densities of the best and worst performing samples based on FENSE scores on EgoSCAPE dense captioning task. We observe that using a single layer, specifically the last, the middle layer or the peak gap layer leads to similar performance improvements as using multiple layers. This shows that the proposed approach can be run on different choice of layers and using a single layer to keep compute minimal. We use the last layer for all our experiments wherever not specified. Also note that we use average attention over all the attention heads of that layer for all the experiments.

\begin{table}[]
\centering
\caption{Effect of different reward weights on the performance of ATTEST framework.}
\label{tab:reward_weight_sweep}
\resizebox{0.85\columnwidth}{!}{%
\begin{tabular}{ccc|cccc|cc}
\hline\hline
 &  &  & \multicolumn{4}{c|}{\textbf{EgoSCAPE Dense Captioning}} & \multicolumn{2}{c}{\textbf{Temporal Grounding}} \\ \cline{4-9} 
 &  &  & \multicolumn{2}{c|}{\textbf{Overall}} & \multicolumn{2}{c|}{\textbf{Long (\textgreater{}120s)}} & \textbf{Clotho-Moment} & \textbf{CASTELLA} \\ \cline{4-9} 
\multirow{-3}{*}{\textbf{$\eta_1 (R_\text{CAG})$}} & \multirow{-3}{*}{\textbf{$\eta_2 (R_\text{AD})$}} & \multirow{-3}{*}{\textbf{$\eta_3 (R_\text{DAC})$}} & \textbf{F@0.5} & \multicolumn{1}{c|}{\textbf{Hal\%}} & \textbf{F@0.5} & \textbf{Hal\%} & \textbf{R@0.5} & \textbf{R@0.5} \\ \hline
\rowcolor{metalight}
\textbf{1} & \textbf{0.2} & \textbf{1} & \textbf{0.427} & \multicolumn{1}{c|}{\textbf{30.5}} & \textbf{0.414} & \textbf{34.2} & \textbf{0.909} & \textbf{0.297} \\
0.25 & {\color[HTML]{9B9B9B} 0.2} & {\color[HTML]{9B9B9B} 1} & 0.415 & \multicolumn{1}{c|}{35.5} & 0.389 & 37.3 & 0.901 & 0.256 \\
0.5 & {\color[HTML]{9B9B9B} 0.2} & {\color[HTML]{9B9B9B} 1} & 0.424 & \multicolumn{1}{c|}{32.3} & 0.406 & 35.1 & 0.905 & 0.285 \\
2 & {\color[HTML]{9B9B9B} 0.2} & {\color[HTML]{9B9B9B} 1} & 0.418 & \multicolumn{1}{c|}{30.8} & 0.401 & 34.5 & 0.906 & 0.267 \\
{\color[HTML]{9B9B9B} 1} & 0.05 & {\color[HTML]{9B9B9B} 1} & 0.414 & \multicolumn{1}{c|}{32.2} & 0.395 & 35.3 & 0.902 & 0.261 \\
{\color[HTML]{9B9B9B} 1} & 0.1 & {\color[HTML]{9B9B9B} 1} & 0.425 & \multicolumn{1}{c|}{31.2} & 0.410 & 34.6 & 0.903 & 0.279 \\
{\color[HTML]{9B9B9B} 1} & 0.5 & {\color[HTML]{9B9B9B} 1} & 0.410 & \multicolumn{1}{c|}{29.5} & 0.389 & 33.8 & 0.891 & 0.250 \\
{\color[HTML]{9B9B9B} 1} & {\color[HTML]{9B9B9B} 0.2} & 0.25 & 0.411 & \multicolumn{1}{c|}{31.0} & 0.391 & 35.0 & 0.893 & 0.251 \\
{\color[HTML]{9B9B9B} 1} & {\color[HTML]{9B9B9B} 0.2} & 0.5 & 0.425 & \multicolumn{1}{c|}{31.6} & 0.399 & 34.9 & 0.901 & 0.281 \\
{\color[HTML]{9B9B9B} 1} & {\color[HTML]{9B9B9B} 0.2} & 2 & 0.426 & \multicolumn{1}{c|}{32.6} & 0.410 & 36.8 & 0.909 & 0.291 \\ \hline\hline
\end{tabular}%
}
\end{table}

\begin{table}[]
\centering
\caption{Effect of varying corruption operator $M(\cdot)$ and fraction $p$ used in $R_\text{CAG}$.}
\label{tab:r_cag_modification_sweep}
\resizebox{\columnwidth}{!}{%
\begin{tabular}{cc|cccc|cc}
\hline\hline
 &  & \multicolumn{4}{c|}{\textbf{EgoSCAPE Dense Captioning}} & \multicolumn{2}{c}{\textbf{Temporal Grounding}} \\ \cline{3-8} 
 &  & \multicolumn{2}{c|}{\textbf{Overall}} & \multicolumn{2}{c|}{\textbf{Long (\textgreater{}120s)}} & \textbf{Clotho-Moment} & \textbf{CASTELLA} \\ \cline{3-8} 
\multirow{-3}{*}{\textbf{\begin{tabular}[c]{@{}c@{}}Corruption\\ Operator (M)\end{tabular}}} & \multirow{-3}{*}{\textbf{\begin{tabular}[c]{@{}c@{}}Corruption\\ Fraction (p)\end{tabular}}} & \textbf{F@0.5} & \multicolumn{1}{c|}{\textbf{Hal\%}} & \textbf{F@0.5} & \textbf{Hal\%} & \textbf{R@0.5} & \textbf{R@0.5} \\ \hline
\rowcolor{metalight}
\textbf{silence} & \textbf{0.3} & \textbf{0.427} & \multicolumn{1}{c|}{\textbf{30.5}} & \textbf{0.414} & \textbf{34.2} & \textbf{0.909} & \textbf{0.297} \\
pink noise & {\color[HTML]{9B9B9B} 0.3} & 0.419 & \multicolumn{1}{c|}{31.2} & 0.404 & 35.6 & 0.902 & 0.265 \\
time shuffle & {\color[HTML]{9B9B9B} 0.3} & 0.421 & \multicolumn{1}{c|}{32.1} & 0.402 & 36.0 & 0.895 & 0.268 \\
audio swap & {\color[HTML]{9B9B9B} 0.3} & 0.411 & \multicolumn{1}{c|}{31.0} & 0.398 & 35.2 & 0.897 & 0.259 \\
low-pass & {\color[HTML]{9B9B9B} 0.3} & 0.409 & \multicolumn{1}{c|}{31.4} & 0.399 & 35.1 & 0.894 & 0.255 \\
{\color[HTML]{9B9B9B} silence} & 0.1 & 0.421 & \multicolumn{1}{c|}{32.8} & 0.403 & 35.9 & 0.900 & 0.269 \\
{\color[HTML]{9B9B9B} silence} & 0.2 & 0.424 & \multicolumn{1}{c|}{31.5} & 0.408 & 35.2 & 0.908 & 0.278 \\
{\color[HTML]{9B9B9B} silence} & 0.4 & 0.422 & \multicolumn{1}{c|}{30.7} & 0.411 & 34.0 & 0.903 & 0.288 \\
{\color[HTML]{9B9B9B} silence} & 0.5 & 0.409 & \multicolumn{1}{c|}{31.9} & 0.400 & 35.3 & 0.898 & 0.271 \\ \hline\hline
\end{tabular}%
}
\end{table}

\begin{table}[b]
\centering
\caption{Effect of using different layers in $R_\text{AD}$. Peak gap represents using the layer where the SFT checkpoint has the most attention density gap between the best and the worst FENSE scoring samples as mentioned in \S\ref{subsec:attention_behavior}.}
\label{tab:r_ad_attention_layer}
\resizebox{\columnwidth}{!}{%
\begin{tabular}{c|cccc|cc}
\hline\hline
\multirow{3}{*}{\textbf{\begin{tabular}[c]{@{}c@{}}Layers used\\ for $R_\text{AD}$\end{tabular}}} & \multicolumn{4}{c|}{\textbf{EgoSCAPE Dense Captioning}} & \multicolumn{2}{c}{\textbf{Temporal Grounding}} \\ \cline{2-7} 
 & \multicolumn{2}{c|}{\textbf{Overall}} & \multicolumn{2}{c|}{\textbf{Long (\textgreater{}120s)}} & \textbf{Clotho-Moment} & \textbf{CASTELLA} \\ \cline{2-7} 
 & \textbf{F@0.5} & \multicolumn{1}{c|}{\textbf{Hal\%}} & \textbf{F@0.5} & \textbf{Hal\%} & \textbf{R@0.5} & \textbf{R@0.5} \\ \hline
 \rowcolor{metalight}
\textbf{Last} & \textbf{0.427} & \multicolumn{1}{c|}{\textbf{30.5}} & \textbf{0.414} & \textbf{34.2} & \textbf{0.909} & \textbf{0.297} \\
First & 0.419 & \multicolumn{1}{c|}{31.6} & 0.408 & 34.9 & 0.905 & 0.289 \\
Middle & 0.426 & \multicolumn{1}{c|}{30.4} & 0.409 & 34.5 & 0.905 & 0.281 \\
First + Middle + Last & 0.427 & \multicolumn{1}{c|}{30.8} & 0.412 & 34.4 & 0.909 & 0.293 \\
Every 3 & 0.424 & \multicolumn{1}{c|}{30.1} & 0.410 & 33.8 & 0.899 & 0.287 \\
Peak Gap & 0.426 & \multicolumn{1}{c|}{30.5} & 0.411 & 34.1 & 0.907 & 0.293 \\ \hline\hline
\end{tabular}%
}
\end{table}

\subsection{Robustness to hallucination}
\label{appsubsec:robustness_halluc_extended}

\begin{wrapfigure}{r}{0.7\textwidth}
    \vspace{-2em}
    \centering
    \includegraphics[width=0.7\textwidth]{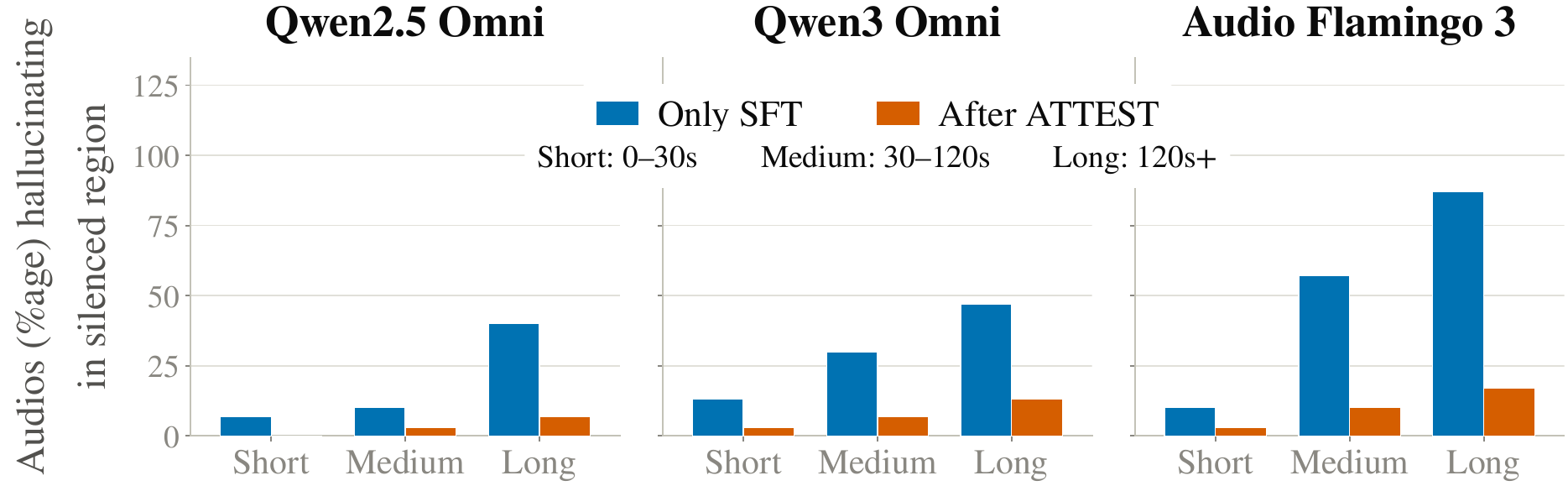}
    \caption{Percentage of audios in EgoSCAPE where at least one dense captioning segment was predicted in a silenced region.}
    \label{fig:robustness_silence_analysis}
    \vspace{-10pt}
\end{wrapfigure}
We perform additional analysis to show the robustness of ATTEST-tuned models to different types of interventions. In \cref{fig:robustness_silence_analysis}, we include silence of random duration between 20-30\% of the audio duration into the EgoSCAPE dense captioning samples and report the fraction of audios in which at least one of the predicted segments had its center in the silenced region, indicating hallucination. We can see that only performing SFT, makes the model prone to predicting hallucinated events in the silenced regions, whereas ATTEST significantly reduces the hallucinations.

\begin{figure}[t]
    \centering
    \begin{subfigure}[b]{0.32\textwidth}
        \centering
        \includegraphics[width=\textwidth]{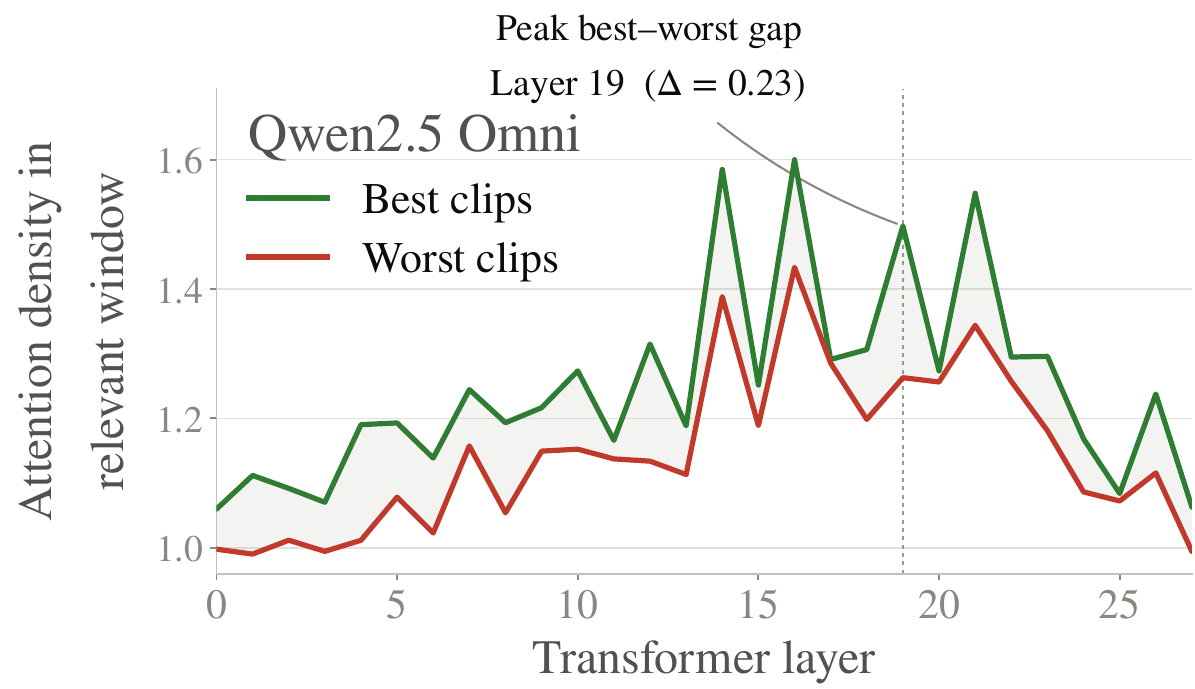}
    \end{subfigure}
    \hfill
    \begin{subfigure}[b]{0.32\textwidth}
        \centering
        \includegraphics[width=\textwidth]{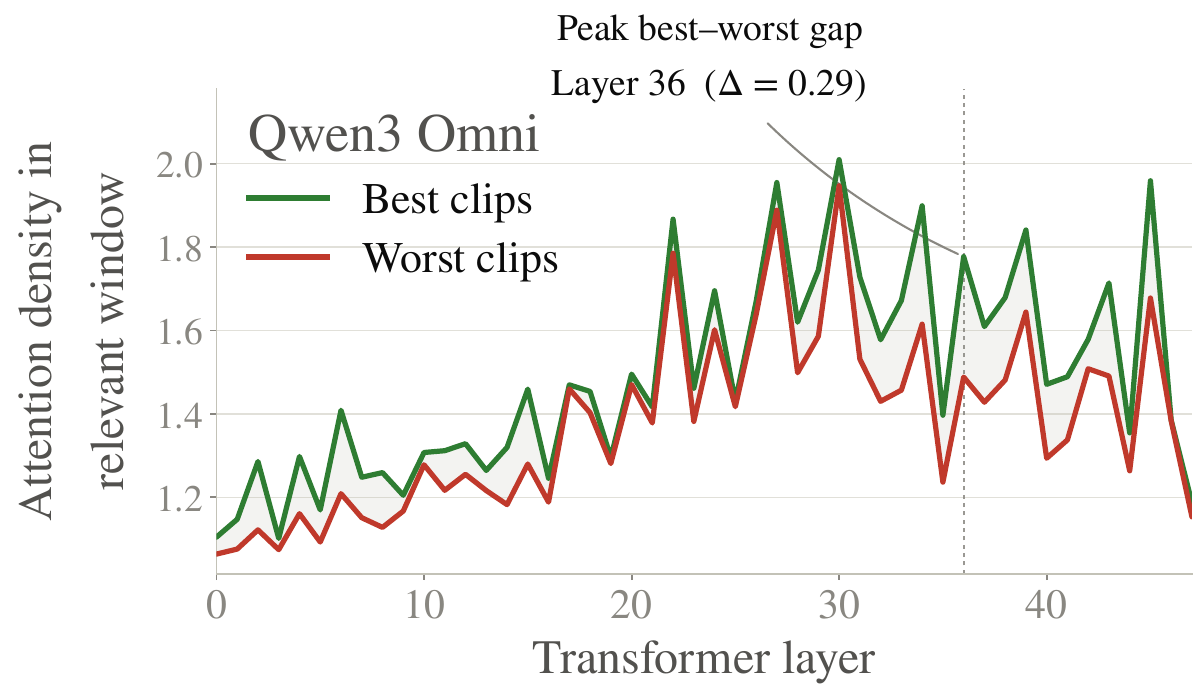}
    \end{subfigure}
    \hfill
    \begin{subfigure}[b]{0.32\textwidth}
        \centering
        \includegraphics[width=\textwidth]{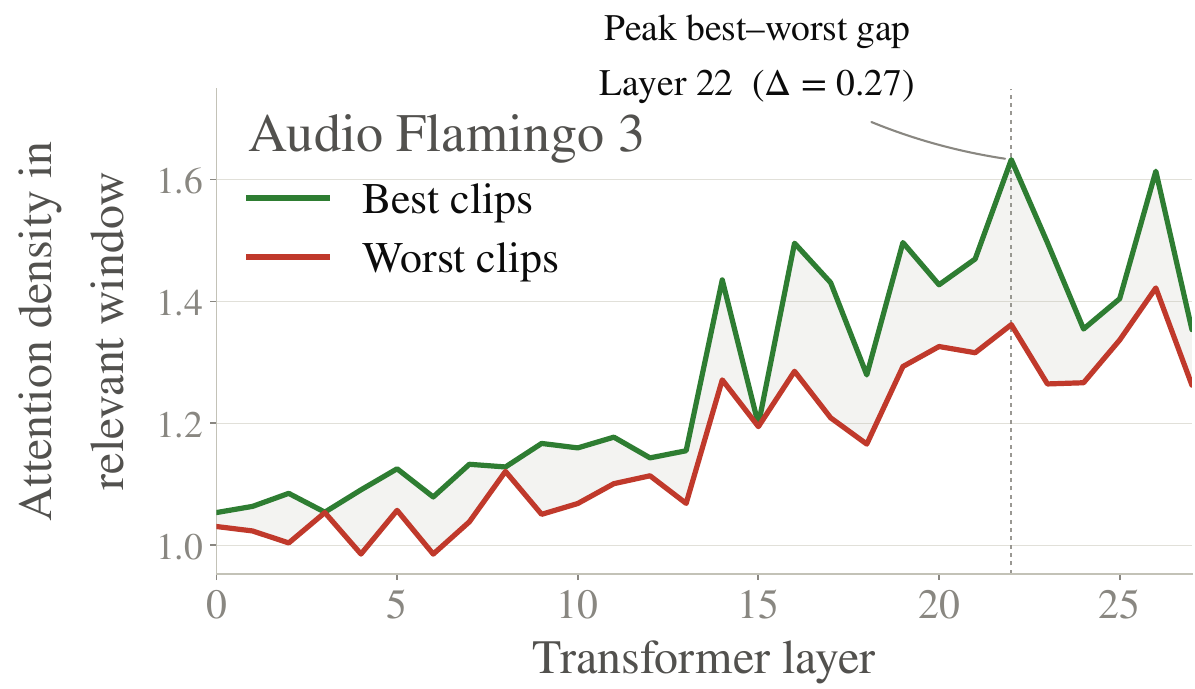}
    \end{subfigure}
    \caption{Attention density over the relevant audio tokens across the layers for the best performing and the worst performing samples in EgoSCAPE after performing SFT on different models.}
    \label{fig:attention_density_best_worst}
\end{figure}

\begin{figure}[t]
    \centering
    \begin{subfigure}[b]{0.32\textwidth}
        \centering
        \includegraphics[width=\textwidth]{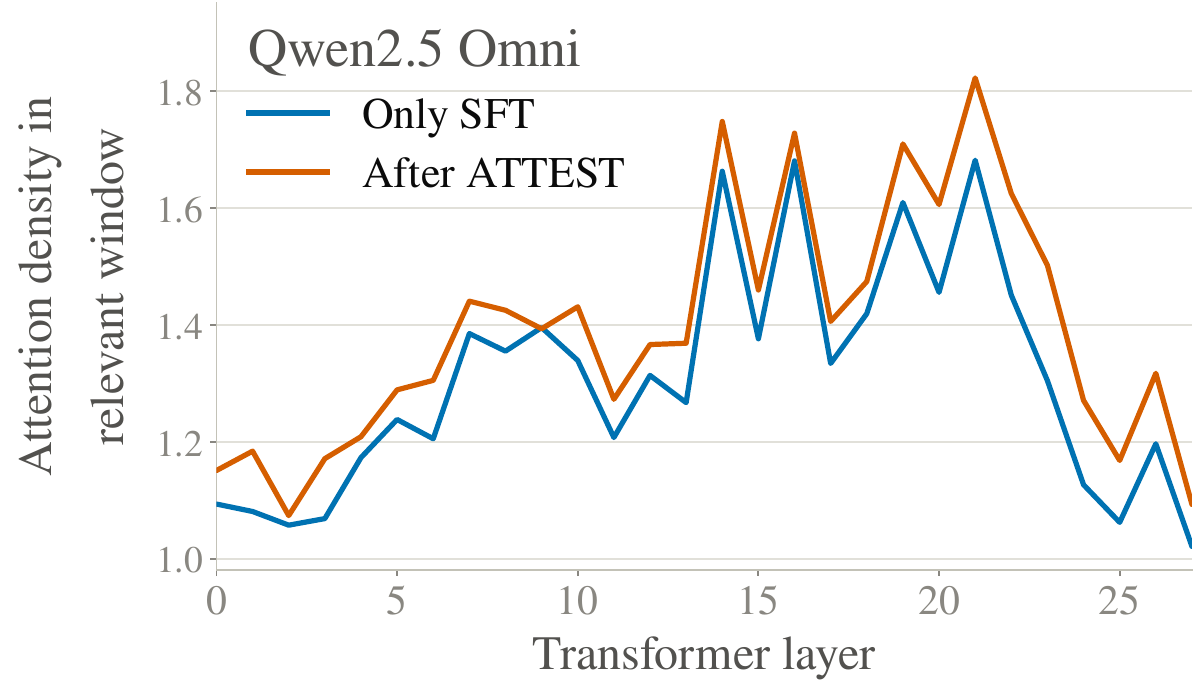}
    \end{subfigure}
    \hfill
    \begin{subfigure}[b]{0.32\textwidth}
        \centering
        \includegraphics[width=\textwidth]{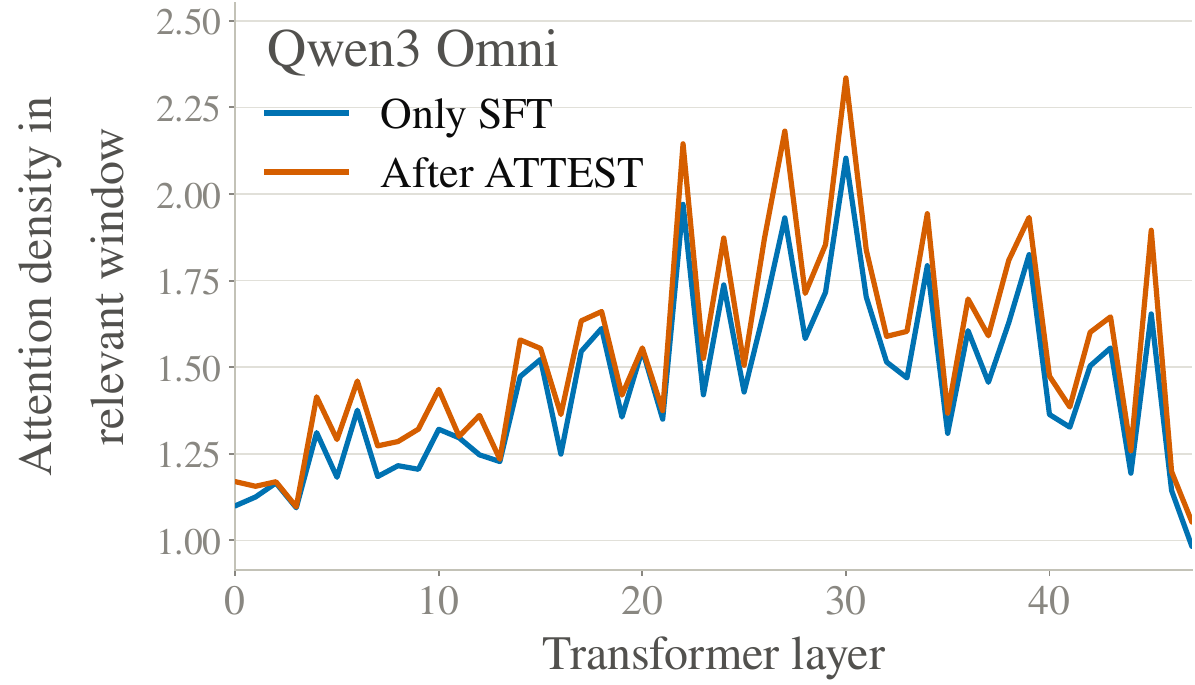}
    \end{subfigure}
    \hfill
    \begin{subfigure}[b]{0.32\textwidth}
        \centering
        \includegraphics[width=\textwidth]{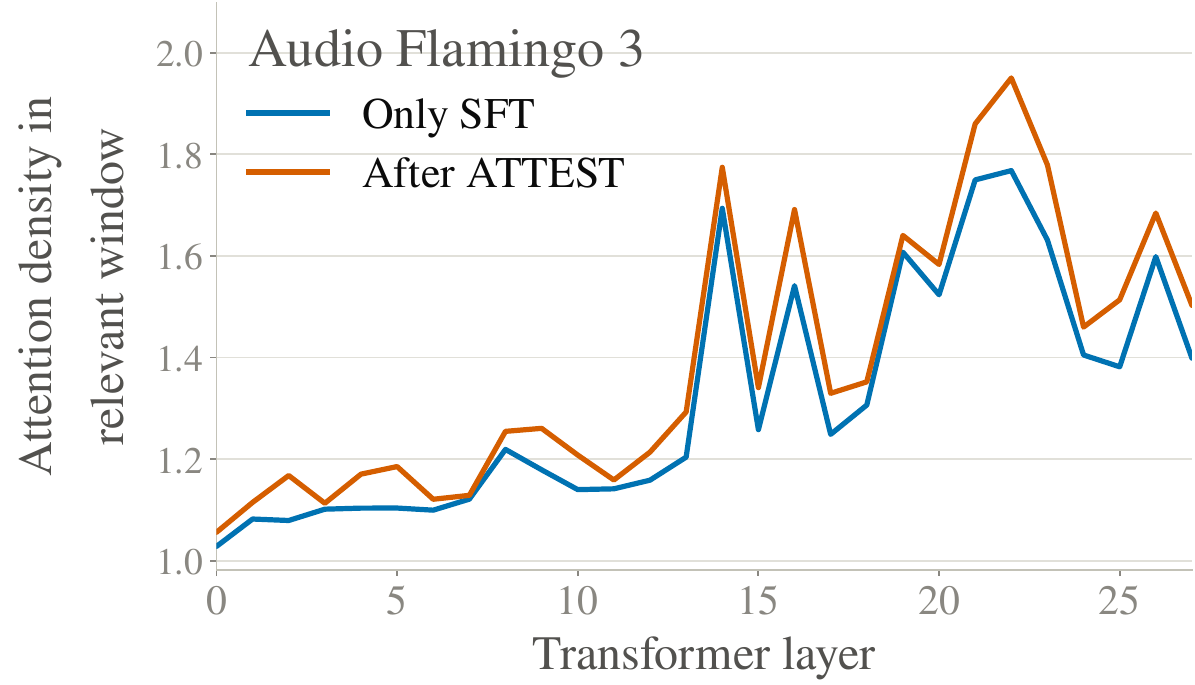}
    \end{subfigure}
    \caption{Attention density across the layers over the relevant audio tokens before (only SFT) and after tuning the model with the ATTEST framework averaged over all the samples of EgoSCAPE dataset.}
    \label{fig:attention_density_before_after_attest}
\end{figure}

\subsection{Attention behavior}
\label{subsec:attention_behavior}

\cref{fig:attention_density_best_worst} shows how the attention pattern differs across the best and the worst scoring samples according to FENSE scores on the EgoSCAPE dense captioning task. We ensure that the number of segments and the duration of samples in the best and worst performing samples are the same to ensure samples are of similar difficulty. We can observe that across the layers, the samples that attend to relevant audio tokens while producing sound event descriptions have better FENSE scores. This observation is the motivation to develop the attention density reward. Moreover, to show the efficacy of $R_\text{AD}$ on correcting model's attention patterns, we show the attention density of different models across layers before and after applying ATTEST in \cref{fig:attention_density_before_after_attest}. We can see that the attention density increases on relevant tokens across all the layers despite the supervision provided only on the last layer.

\clearpage

\subsection{Importance of real data}
\label{appsubsec:real_data_importance}

\begin{wraptable}{r}{0.6\columnwidth}
\centering
\vspace{-\intextsep}
\caption{Dense captioning results on using Real vs Synthetic long audio training data on EgoSCAPE as well as AudioSet-SL (by joining multiple clips together).}
\label{tab:data_ablation}
\resizebox{0.6\columnwidth}{!}{%
\begin{tabular}{l|cc|cc}
\hline\hline
\multicolumn{1}{c|}{\multirow{2}{*}{\textbf{Training Data}}} & \multicolumn{2}{c|}{\textbf{EgoSCAPE ($>$120s)}} & \multicolumn{2}{c}{\textbf{AudioSet-SL (Joined-Long)}} \\ \cline{2-5}
\multicolumn{1}{c|}{} & \textbf{F@0.5} & \textbf{Hal\%} & \textbf{F@0.5} & \textbf{Hal\%} \\ \hline
\rowcolor{metalight}
\textbf{Real} & \textbf{0.425} & \textbf{33.2} & 0.398 & 36.2 \\
Synthetic & 0.361 & 43.8 & \textbf{0.404} & \textbf{34.6} \\
Real + Synthetic & 0.416 & 34.9 & 0.399 & 35.3 \\ \hline\hline
\end{tabular}%
}
\end{wraptable}
We can create long-audios and corresponding temporal annotations for such long audios by joining short audios from a dataset (e.g. AudioSet-SL) which already has temporal annotations. To show that such a combination does not capture the real data distribution, we report the results of training the Qwen2.5 Omni model using real and/or synthetic data using the ATTEST framework. The synthetic data is constructed by joining multiple AudioSet-SL audios together randomly and creating temporal annotations for them by off-setting the original timestamps in AudioSet-SL. For reporting performance, we use a source-stripped version of EgoSCAPE which has the dense captions without the sound sources for fair comparison with AudioSet-SL. We can clearly observe that the model's performance on EgoSCAPE on using synthetic data only is significantly worse than using Real data. Moreover, we can observe that the model's performance on EgoSCAPE is better compared to synthetically joined AudioSet-SL because real data has continuous audio and continuous context makes the dense captioning task easier.

\clearpage

\section{Prompts}
\label{appsec:prompts}

\begingroup
\footnotesize
\setlength{\extrarowheight}{2pt}
\begin{longtable}{@{}p{\textwidth}@{}}
\caption{Prompt used to verify and refine the sound events detected by the sound event detector model as described in \S\ref{sec:egoscape_data}.}
\label{tab:prompt_sed_verify}\\
\toprule
\endfirsthead
\toprule
\endhead
\midrule \multicolumn{1}{r}{\emph{continued on next page}}\\
\endfoot
\bottomrule
\endlastfoot
\textbf{Task.} You are annotating a short audio clip (up to 30 seconds) from an egocentric recording. For each candidate sound event detected in the clip, decide keep / drop / relabel by listening to the audio and watching the video. \\ \addlinespace
\textbf{Inputs.} You are given: \newline
1.~a short video clip (typically up to 30 seconds) with synchronized audio --- watch and listen to it; \newline
2.~a list of candidate sound events, each with an idx, a class name, and onset/offset times in seconds (relative to the clip start at 0.0); \newline
3.~sometimes a NARRATION section (see below). \\ \addlinespace
\textbf{Guidelines.} \newline
\emph{Decisions --- for each candidate choose one:} \newline
\textbullet~\textbf{keep} --- the event is present in the audio at roughly the indicated time, and its class label is correct. You can optionally refine the sound event class to make it more accurate. \newline
\textbullet~\textbf{drop} --- the event is not actually present. \newline
\textbullet~\textbf{relabel} --- a sound is audible at this time but the class name is wrong; the visual frames at the same timestamp support a different class. Use relabel whenever the class is wrong but a sound is present --- it is usually a better answer than drop in that case. \newline
\emph{How to decide:} \newline
\textbullet~The candidate list comes from a high-recall audio model. Most events are real, but some are false positives, wrong-class, or duplicates. \newline
\textbullet~Listen to the audio at [onset, offset] before deciding. Faint, brief, distant, or background sounds count as present --- do NOT drop just because a sound is subtle, as long as you can hear something matching the class. \newline
\textbullet~Drop when: (a) the audio at [onset, offset] is genuinely silent or contains no sound matching the class; (b) the candidate is an obvious duplicate of another candidate of the same class with near-identical timing AND the audio clearly contains only one such event. \newline
\textbullet~Relabel when: the audio has a clear sound at [onset, offset], but it's not what the class says --- and the video frames disambiguate the true class. \newline
\textbullet~Keep when: you can hear a sound matching the class, even if faint or partially obscured. \newline
\emph{Rules:} \newline
\textbullet~Do not add new events that were not in the candidate list. \newline
\textbullet~Do not change onset or offset. \newline
\textbullet~Every candidate (every idx in the input) must appear in decisions exactly once. \newline
\textbullet~If two candidates of the same class overlap heavily and both look correct, keep both. \\ \addlinespace
\textbf{Using the narration.} A NARRATION section may list what people DID and SAID during the clip (window-relative seconds), split into the WEARER (\#C = the FIRST PERSON / camera wearer) and OTHER people (\#O/\#X, sometimes named). Use it as a STRONG HINT: \newline
\textbullet~\#C or \#O actions that make sound (handling objects, tools, walking, pouring) support KEEPING a co-occurring candidate of the matching class. \newline
\textbullet~When a candidate's class conflicts with what the narration says is happening at that time, prefer RELABEL (if the video supports the true class) or DROP. \newline
\textbullet~Narration is sparse and may miss events or be slightly off in time; OVERRIDE it when what you actually hear/see clearly conflicts. When no narration is provided, judge from audio+video alone. \\ \addlinespace
\textbf{Output format.} Return STRICT JSON, in a single \texttt{json} fenced block, with this exact schema: \newline
\texttt{\{"decisions": [\{"idx": \textless int\textgreater, "action": "keep", "refined\_class": "\textless refined class label\textgreater", "reason": "\textless short reason\textgreater"\}, \{"idx": \textless int\textgreater, "action": "drop", "reason": "\textless short reason\textgreater"\}, \{"idx": \textless int\textgreater, "action": "relabel", "new\_class": "\textless short class label\textgreater", "reason": "\textless short reason\textgreater"\}]\}} \newline
Reasons should be one short sentence, no more than 120 characters. No prose outside the JSON block. \\ \addlinespace
\textbf{User-turn template} (the clip's video+audio is attached): \newline
\texttt{Clip: \{chunk\_label\}~~(\{duration\_sec:.2f\} s, video + audio)} \par\vspace{2pt}
\emph{Candidate events (sorted by onset), as a \texttt{json} block --- one entry per SED candidate:} \newline
\texttt{[\{"idx": \textless int\textgreater, "class": "\textless class name\textgreater", "onset": \textless float\textgreater, "offset": \textless float\textgreater\}, \ldots]} \par\vspace{2pt}
\emph{NARRATION (what people did/said; window-relative seconds; \#C = the wearer/first person, others are \#O/\#X):} \newline
\emph{~~WEARER / first person (\#C) --- actions \& speech:} \newline
\texttt{~~~~\{t:.1f\}~~\{text\}} \newline
\emph{~~OTHER people (\#O/\#X, may be named) --- actions \& speech:} \newline
\texttt{~~~~\{t:.1f\}~~[\#\{tag\}] \{text\}} \par\vspace{2pt}
\texttt{Return the JSON block only.} \\
\end{longtable}
\endgroup

\begingroup
\footnotesize
\setlength{\extrarowheight}{2pt}
\begin{longtable}{@{}p{\textwidth}@{}}
\caption{Prompt used to generate the audio timeline events using the dense audio caption sound events as described in \S\ref{sec:egoscape_data}.}
\label{tab:prompt_audio_timeline_generation}\\
\toprule
\endfirsthead
\toprule
\endhead
\midrule \multicolumn{1}{r}{\emph{continued on next page}}\\
\endfoot
\bottomrule
\endlastfoot
\textbf{Task.} You are annotating a short audio clip (up to 30 seconds) from an egocentric recording. Write a SMALL NUMBER of coarse, scene-level captions with timestamps that summarize what is happening in the soundscape; together they should cover almost the entire clip except stretches of genuine silence. You HEAR the audio and can SEE the video, but the video is only for disambiguating what you hear. \\ \addlinespace
\textbf{Inputs.} You are given: \newline
1.~the audio clip (video available as a disambiguation aid) --- watch and listen to it; \newline
2.~a list of verified sound events with timestamps (from an SED model, model-audited; sorted by onset); \newline
3.~any transcribed speech (with lang and, when non-English, a text\_en translation); \newline
4.~sometimes a NARRATION section (see below). \\ \addlinespace
\textbf{Guidelines.} \newline
\emph{Scene grouping is the primary task:} \newline
\textbullet~Treat overlapping or interleaved fine-grained events as ONE layered scene. Name the dominant sound source; mention background/co-occurring sounds inside the SAME caption. \newline
\textbullet~Repeated instances of the same sound within a short window (repeated clinks, taps, footsteps) belong in ONE caption describing the pattern, NOT one caption per instance. \newline
\textbullet~A new caption is only warranted when the scene MEANINGFULLY changes: the dominant sound source changes, a new sonic environment begins, or a distinct isolated event punctuates the ambience. \newline
\emph{Worked example} (what ``scene grouping'' means): given events on 0--10\,s (coughs at 1.1--2.3 and 4.1--4.7; dishwasher running 0.0--5.6; dish clanks at 0.8--1.2 and 4.5--5.0), WRONG is one caption per event; CORRECT is a single caption: ``0.0 - 5.6, Dishes clank inside a running dishwasher while a man coughs briefly in the foreground.'' \\ \addlinespace
\emph{Audio is the subject; video is only a disambiguation aid (strict):} \newline
\textbullet~EVERY caption MUST describe a sound that is actually AUDIBLE. If something is visible but produces NO audible sound, DO NOT mention it at all. \newline
\textbullet~The video is allowed only to decide the SOURCE of a sound you actually hear (e.g.\ a metallic clang + a visible dishwasher rack $\rightarrow$ ``dishes clank''). You may NOT add silent actions. \newline
\textbullet~Examples of what NOT to write: ``The person applies yellow wood glue from a bottle'' (visual, no distinctive sound); ``A woman gestures toward the camera'' (silent); ``The person walks toward the kitchen'' (unless footsteps are audible). \newline
\emph{Speech content must be captured (when English or when translated):} \newline
\textbullet~If English speech is audible (or the transcript's lang is en), include a short verbatim quote (\textless= 15 words, in double quotes) OR a one-sentence summary, inside the caption covering that timespan. Do NOT just write ``a person speaks''. \newline
\textbullet~If the speech is non-English, name the language and, if text\_en is provided, include a short summary of the translated content; otherwise describe manner and prosody. \newline
\textbullet~The speech description belongs INSIDE the same layered caption, not as its own separate caption (unless speech is the ONLY thing happening for a long stretch). \newline
\emph{Timestamps must be audio-anchored, not grid-snapped:} \newline
\textbullet~Anchor caption boundaries to the ACTUAL audible transitions; the verified event list gives the dense onset/offset timestamps --- USE THEM. \newline
\textbullet~Do NOT snap boundaries to multiples of 5 seconds. Prefer real audio-anchored values like 3.2, 11.7, 24.6. Timestamps are float seconds relative to clip start (0.0), one decimal. \newline
\emph{Content style and coverage:} \newline
\textbullet~Describe what the sound ACTUALLY SOUNDS LIKE (pitch, texture, rhythm, loudness, duration, timbre, spatial impression) --- not just the activity. Distinguish foreground from background inside a single caption. Aim for 15--30 words per caption. \newline
\textbullet~The union of your caption spans should cover the audible portion of the clip; consecutive spans should be contiguous when scenes flow into each other; overlapping spans only when two independent layers cannot be merged. \\ \addlinespace
\textbf{Using the narration.} A NARRATION section may list what people DID and SAID (window-relative seconds), split into the WEARER (\#C = the FIRST PERSON) and OTHER people (\#O/\#X, sometimes named). Use it as a STRONG HINT for WHO is doing/saying what: \newline
\textbullet~\#C actions (chopping, pouring, washing, typing) and \#O actions help you name the correct sound SOURCE inside a caption; use the names to keep distinct people apart. \newline
\textbullet~When \#C or an \#O person is narrated talking around a speech segment, attribute the speech accordingly and capture its content in the caption covering that span. \newline
\textbullet~Narration describes VISIBLE actions/speech, NOT sounds --- mention a narrated action only if it is actually AUDIBLE. OVERRIDE it when what you hear/see conflicts; when none is provided, judge from audio+video alone. \\ \addlinespace
\textbf{Output format.} STRICT JSON in a single \texttt{json} fenced block: \texttt{\{"captions": [\{"start": 0.0, "end": 8.0, "text": "..."\}]\}}. No prose outside the JSON block. \\ \addlinespace
\textbf{User-turn template} (video attached as disambiguation aid): \newline
\texttt{Clip: \{clip\_label\}~~(\{duration\_sec:.2f\} s, audio with video available as disambiguation aid)} \par\vspace{2pt}
\emph{Verified sound events (from an SED model, model-audited; sorted by onset), as a \texttt{json} block:} \newline
\texttt{[\{"class": "\textless name\textgreater", "onset": \textless float\textgreater, "offset": \textless float\textgreater\}, \ldots]} \par\vspace{2pt}
\emph{Speech transcription (included only when speech is present), as a \texttt{json} block:} \newline
\texttt{[\{"start": \textless float\textgreater, "end": \textless float\textgreater, "lang": "\textless lang\textgreater", "text\_en": "\textless english\textgreater"\}, \ldots]} \par\vspace{2pt}
\emph{NARRATION (what people did/said; window-relative seconds; \#C = the wearer/first person, others are \#O/\#X):} \newline
\emph{~~WEARER / first person (\#C) --- actions \& speech:} \newline
\texttt{~~~~\{t:.1f\}~~\{text\}} \newline
\emph{~~OTHER people (\#O/\#X, may be named) --- actions \& speech:} \newline
\texttt{~~~~\{t:.1f\}~~[\#\{tag\}] \{text\}} \par\vspace{2pt}
\texttt{Return the JSON block only.} \\
\end{longtable}
\endgroup

\begingroup
\footnotesize
\setlength{\extrarowheight}{2pt}
\begin{longtable}{@{}p{\textwidth}@{}}
\caption{Prompt used to tag different acoustic events in dense audio captioning to their sound sources as mentioned in \S\ref{sec:egoscape_data}. The injected \texttt{\{taxonomy\}} defines seven source keys: \texttt{self\_body},
\texttt{self\_speech}, \texttt{first\_person\_interaction},
\texttt{environment\_ambient}, \texttt{electronic\_media}, \texttt{other\_speech},
\texttt{uncertain}.}
\label{tab:prompt_densesource}\\
\toprule
\endfirsthead
\toprule
\endhead
\midrule \multicolumn{1}{r}{\emph{continued on next page}}\\
\endfoot
\bottomrule
\endlastfoot
\textbf{Task.} You classify the SOURCE of sounds in an egocentric (head-worn camera) video clip. The person wearing the camera is the `wearer' / FIRST PERSON --- you never see their face; you see the scene from their eyes and their hands. For EACH segment, watch what happens during its time window and listen, then assign its PRIMARY source class from the taxonomy (you decide how many apply --- usually exactly one). \\ \addlinespace
\textbf{Inputs.} You are given: \newline
1.~the CLIP itself (video with its audio) --- watch and listen to it; \newline
2.~a numbered list of SOUND SEGMENTS already detected in the clip (id, [start - end] seconds, and a short description); \newline
3.~sometimes a NARRATION section (see below). \\ \addlinespace
\textbf{Guidelines.} \newline
\emph{Taxonomy (injected):} \newline
\texttt{\{taxonomy\}} \newline
\emph{Decide with this procedure for every segment:} \newline
1.~Is the sound SPEECH / a VOICE? \newline
\quad -- Yes, and it is the WEARER'S voice (usually the LOUDEST, closest, most `present' voice, with NO person visible in frame whose mouth moves in sync) $\rightarrow$ self\_speech. \newline
\quad -- Yes, and it is ANOTHER co-present person $\rightarrow$ other\_speech. \newline
\quad -- Yes, but it comes from a screen/speaker/recording $\rightarrow$ electronic\_media. \newline
2.~Not speech? Then: \newline
\quad -- the wearer's own BODY (breathing, footsteps, coughing, chewing) $\rightarrow$ self\_body. \newline
\quad -- the wearer ACTING on objects/environment (handling, tools, tap, typing, pouring, washing) $\rightarrow$ first\_person\_interaction. \newline
\quad -- a device or media $\rightarrow$ electronic\_media. \newline
\quad -- anything else $\rightarrow$ environment\_ambient. \newline
The same description can recur at several segments; judge each on its own window. Use `uncertain' only when you truly cannot decide. \\ \addlinespace
\textbf{Using the narration.} A NARRATION section may list what people DID and SAID during the clip (window-relative seconds), split into the WEARER (\#C = the FIRST PERSON / camera wearer --- that is YOU/the viewpoint) and OTHER people (\#O/\#X, sometimes named like `woman H'). Narration describes VISIBLE actions/speech, NOT sounds --- use it as a STRONG HINT for WHO is doing/saying what: \newline
\textbullet~When the WEARER (\#C) is narrated talking around a speech segment $\rightarrow$ that speech is self\_speech. When an OTHER person is narrated talking $\rightarrow$ other\_speech (use the names to keep distinct people apart). \newline
\textbullet~\#C actions (kneading, picking, chopping, pouring, washing) support first\_person\_interaction for co-occurring object sounds; the wearer walking/breathing $\rightarrow$ self\_body. \newline
\textbullet~\#O actions (another person handling things, walking) support environment\_ambient for their non-speech sounds; another person's speech $\rightarrow$ other\_speech. \newline
\textbullet~Narration is sparse and may miss events or be slightly off in time; OVERRIDE it when what you actually hear/see clearly conflicts. When no narration is provided, judge by the audio/video alone. \\ \addlinespace
\textbf{Output format.} STRICT JSON only, in one \texttt{json} fenced block, with one object for EVERY segment id given (same ids, no more, no fewer): \newline
\texttt{\{\{ \newline
\quad "segments": [ \newline
\quad\quad \{\{ \newline
\quad\quad\quad "seg\_id": "\textless the exact id\textgreater", \newline
\quad\quad\quad "source\_primary": ["\textless taxonomy key(s)\textgreater"], \newline
\quad\quad\quad "source\_labels": ["\textless same as primary, plus any extra overlapping source\textgreater"], \newline
\quad\quad\quad "confidence": \textless 0.0-1.0\textgreater, \newline
\quad\quad\quad "rationale": "\textless one short clause; cite the narrated actor if you used narration\textgreater" \newline
\quad\quad \}\} \newline
\quad ] \newline
\}\}} \\ \addlinespace
\textbf{User-turn template} (the clip's video+audio is attached): \newline
\texttt{CLIP DURATION: \{duration\_sec:.1f\} seconds (the video with audio is attached).} \par\vspace{2pt}
\emph{SOUND SEGMENTS (\{n\} total) --- classify the source of each:} \newline
\texttt{\{seg\_id\}~~[\{start:.1f\} - \{end:.1f\}]~~\{description\}} \newline
\emph{\ldots one line per segment \ldots} \par\vspace{2pt}
\emph{NARRATION (what people did/said; window-relative seconds; \#C = the wearer/first person, others are \#O/\#X):} \newline
\emph{~~WEARER / first person (\#C) --- actions \& speech:} \newline
\texttt{~~~~\{t:.1f\}~~\{text\}} \newline
\emph{~~OTHER people (\#O/\#X, may be named) --- actions \& speech:} \newline
\texttt{~~~~\{t:.1f\}~~[\#\{tag\}] \{text\}} \par\vspace{2pt}
\texttt{Watch and listen to the attached clip, then return the JSON with one object per seg\_id above. Return only the JSON block.} \\
\end{longtable}
\endgroup

\begingroup
\footnotesize
\setlength{\extrarowheight}{2pt}
\begin{longtable}{@{}p{\textwidth}@{}}
\caption{Prompt used to rewrite the audio timeline when merging multiple individually generated chunk audio timelines. }
\label{tab:prompt_timeline_rewrite}\\
\toprule
\endfirsthead
\toprule
\endhead
\midrule \multicolumn{1}{r}{\emph{continued on next page}}\\
\endfoot
\bottomrule
\endlastfoot
\textbf{Task.} You are cleaning up a list of audio captions that has been assembled from one or more 30-second source clips into a single ordered timeline. The listener will read your output start-to-finish as a single audio sample. Remove or rephrase any wording that assumes a different frame of reference (the original 30\,s clip's start/end, or a per-clip event count). \\ \addlinespace
\textbf{Inputs.} You are given: \newline
1.~The ordered caption list of one sample (text only; no audio, no source-chunk metadata), each caption with start, end, and text. \\ \addlinespace
\textbf{Guidelines.} \newline
\emph{Strict rules:} \newline
1.~Do NOT change timestamps. Every caption's start and end must be identical to the input, to at least three decimal places. If you feel you need to change a timestamp, LEAVE THE CAPTION UNCHANGED. \newline
2.~Do NOT change the acoustic description (sound source, character, texture, layering) beyond removing stale positional wording. \newline
3.~Do NOT change speech content: preserve verbatim quotes (in double quotes) and one-sentence summaries of what was said. \newline
4.~Do NOT invent new information, sounds, or speakers. If a phrase is genuinely descriptive (e.g.\ ``the tapping resumes after a brief pause''), keep it. \newline
5.~MERGE two captions if their context is seemingly the same and update their start and end times.
\emph{What to rewrite --- remove or rephrase:} \newline
\textbullet~Position-relative to a chunk boundary: ``towards the end'', ``at the very end'', ``just before the clip ends'', ``at the very beginning'', ``at the start of the clip''. \newline
\textbullet~Cardinal/ordinal counters that referenced the source clip's event count: ``for the first time'', ``for the second time'', ``the third instance'', ``for the last time'', ``once again'', ``yet again'', ``again'' (when it counts events not sounds), ``previously in this clip'', ``earlier in the clip'', ``later in the clip''. \newline
\textbullet~Ordinal digits or short forms that count within the clip: ``1st'', ``2nd'', ``3rd'', ``the third X''. \newline
\emph{Rephrase strategy:} replace with a neutral equivalent when possible (``for the first time'' $\rightarrow$ drop or ``a''; ``towards the end'' $\rightarrow$ drop or a specific timestamp already carried in the timestamp field). If the phrase is only cosmetic, drop it and keep the sentence natural. If removing it would break the sentence, do the minimum rewrite needed to keep the acoustic meaning. If a caption has NONE of these patterns, return it byte-for-byte identical. \\ \addlinespace
\textbf{Output format.} STRICT JSON in a single \texttt{json} fenced block, one item per input caption in the same order, with fields start, end, text (unchanged start/end): \texttt{\{"captions": [\{"start": 0.0, "end": 12.4, "text": "..."\}]\}}. No prose outside the JSON block. No trailing commentary. \\ \addlinespace
\textbf{User-turn template} (text-only): \newline
\texttt{Sample id: \{sample\_id\}} \newline
\texttt{Total sample duration: \{duration:.2f\} s} \newline
\texttt{Number of captions: \{n\}} \par\vspace{2pt}
\emph{Captions (ordered, timeline-local seconds), as a \texttt{json} block:} \newline
\texttt{[\{"start": \textless float\textgreater, "end": \textless float\textgreater, "text": "\textless caption text\textgreater"\}, \ldots]} \par\vspace{2pt}
\texttt{Rewrite each caption per the rules above. Preserve timestamps EXACTLY. Return the JSON block only.} \\
\end{longtable}
\endgroup

\begingroup
\footnotesize
\setlength{\extrarowheight}{2pt}
\begin{longtable}{@{}p{\textwidth}@{}}
\caption{Prompt used to generate fine-grained audio queries for temporal grounding using dense audio captions. We use $K = 2$.}
\label{tab:prompt_tag_fine_grained}\\
\toprule
\endfirsthead
\toprule
\endhead
\midrule \multicolumn{1}{r}{\emph{continued on next page}}\\
\endfoot
\bottomrule
\endlastfoot
\textbf{Task.} You are annotating temporal audio grounding (TAG) data. Given the audio and a text annotation of sound events, produce \{K\} temporal audio grounding QAs that locate specific sounds in time, or abstain if the annotation clearly mismatches the audio. \\ \addlinespace
\textbf{Inputs.} You are given: \newline
1.~the audio recording --- listen carefully; \newline
2.~a text annotation of DETECTED EVENTS (label $\rightarrow$ list of intervals) produced by an earlier automated model, which may be imperfect. \\ \addlinespace
\textbf{Guidelines.} \newline
\emph{Procedure:} (1) LISTEN to the audio; (2) READ the annotation; (3) VERIFY whether it plausibly describes what you hear; (4) if clearly wrong or mismatched, ABSTAIN by returning an empty JSON array []; (5) otherwise produce \{K\} QAs, cross-checking each interval against the audio. \newline
\emph{Merge recurrences across labels:} the annotation may split the SAME underlying sound across multiple event classes (e.g.\ ``Rub'' and ``Scrape''; ``Speech'' and ``Male speech''). Gather intervals from EVERY label that describes the SAME acoustic phenomenon. \newline
\emph{Do NOT invent events:} every query MUST correspond to at least one event class actually listed in the DETECTED EVENTS block. Do not add sounds absent from the annotation, even if audible. \newline
\emph{Avoid near-full-duration intervals:} any single interval whose length exceeds 80\% of \{duration\} is FORBIDDEN unless the source annotation itself lists an interval that long AND no shorter grouping is possible. Prefer tight, specific intervals aligned with actual bursts you hear. \newline
\emph{Requirements for each QA:} \newline
\textbullet~query: a short natural-language phrase (3--12 words) describing a specific perceivable sound the way a human listener would (``scratchy sanding sound on wood'', ``a man speaking'', ``sharp metallic clink''). Not a class name. \newline
\textbullet~intervals: list of [start, end] pairs (seconds, 1-decimal) covering EVERY interval where that sound occurs, as you HEAR them. Timestamps within [0.0, \{duration\}]. No interval may exceed 80\% of \{duration\}. \newline
\textbullet~n\_occurrences\_hint: ``multi'' if intervals has 2+ entries, ``single'' if exactly 1. \newline
\textbullet~verified: true --- on every emitted QA, to confirm you listened and cross-checked. \newline
\emph{Verification, balance, abstain:} seek to each proposed interval and confirm the sound is audible there; drop/trim accordingly. Aim for roughly half the queries with 2+ intervals and half with exactly 1. Return [] if the audio is silent/unintelligible while the annotation describes clear events, if the annotation describes events (e.g.\ English speech) not heard at all, or if annotation and audio disagree on more than roughly half of the described events. \\ \addlinespace
\textbf{Output format.} Return ONLY a JSON array of \{K\} objects (or [] to abstain). No prose, no markdown fences. Example: \texttt{[\{"query": "a scratchy sanding sound on wood", "intervals": [[0.8, 4.7],[12.1, 18.3]], "n\_occurrences\_hint": "multi", "verified": true\}, \{"query": "faint background rustling", "intervals": [[0.0, 30.0]], "n\_occurrences\_hint": "single", "verified": true\}]}. \\ \addlinespace
\textbf{User-turn template} (audio attached; appended below the instructions above): \newline
\texttt{AUDIO CLIP: (attached --- please listen)} \newline
\texttt{duration\_sec: \{duration\}} \par\vspace{2pt}
\emph{DETECTED EVENTS (label $\rightarrow$ list of intervals, from an automated model --- verify against the audio):} \newline
\texttt{\{events\_block\}} --- one line per label: \texttt{- \textless label\textgreater: [\textless on\textgreater, \textless off\textgreater], [\textless on\textgreater, \textless off\textgreater], \ldots} \par\vspace{2pt}
\texttt{Now produce the JSON array with exactly \{K\} queries, or [] if you must abstain.} \\
\end{longtable}
\endgroup

\begingroup
\footnotesize
\setlength{\extrarowheight}{2pt}
\begin{longtable}{@{}p{\textwidth}@{}}
\caption{Prompt used to generate coarse audio queries for temporal grounding using audio timeline events. We use $K = 2$.}
\label{tab:prompt_tag_coarse}\\
\toprule
\endfirsthead
\toprule
\endhead
\midrule \multicolumn{1}{r}{\emph{continued on next page}}\\
\endfoot
\bottomrule
\endlastfoot
\textbf{Task.} You are annotating temporal audio grounding (TAG) data. Given the audio and a coarse, time-aligned narrative, produce \{K\} temporal audio grounding QAs that locate specific sounds in time, or abstain if the narrative clearly mismatches the audio. \\ \addlinespace
\textbf{Inputs.} You are given: \newline
1.~the audio recording --- listen carefully; \newline
2.~a coarse, time-aligned narrative (COARSE CAPTIONS) produced by an earlier model, which may be imperfect or hallucinated. \\ \addlinespace
\textbf{Guidelines.} \newline
\emph{Procedure:} (1) LISTEN carefully; (2) READ the coarse captions; (3) VERIFY whether they plausibly describe what you hear; (4) if clearly wrong or mismatched, ABSTAIN by returning []; (5) otherwise produce \{K\} QAs, cross-checking each interval against the audio. \newline
\emph{Query style --- specific, descriptive (critical):} every query must read like a descriptive acoustic annotation sentence, NOT a terse class-like tag. Include the sound's texture, loudness, rhythm, and prominence, using what you HEAR. Good: ``Gentle, calming music can be faintly heard in the background''; ``A steady, coarse scraping of sandpaper on wood dominates the foreground''; ``Sharp, sudden metallic clanks interrupt the ambience at irregular intervals''. Bad (do NOT produce): ``a woman laughing''; ``sanding sound''; ``background talking''; ``car engine idling''. \newline
\emph{Extract specific sounds and find every occurrence:} captions summarize whole scenes, so a single caption may contain multiple distinct sounds, and the SAME sound may recur across captions worded differently. Decompose the captions into specific perceivable sounds; collect intervals from EVERY caption whose scene contains that sound. \newline
\emph{Do NOT invent events:} every query MUST correspond to a sound explicitly mentioned or unambiguously implied in the COARSE CAPTIONS block. Do not add sounds absent from the captions, even if audible; the purpose is to ground and refine the captions, not extend them. \newline
\emph{Avoid near-full-duration intervals:} any single interval whose length exceeds 80\% of \{duration\} is FORBIDDEN unless the underlying caption span itself is that long AND no tighter interval fits. \newline
\emph{Requirements for each QA:} \newline
\textbullet~query: a descriptive sentence (8--20 words) about ONE specific sound with acoustic modifiers (soft/loud/faint, rhythmic/irregular/continuous, sharp/gentle/muffled, foreground/background). NOT a class name or terse tag. \newline
\textbullet~intervals: list of [start, end] pairs (seconds, 1-decimal), TIGHT to the actual span you hear; no single interval may exceed 80\% of \{duration\}; timestamps within [0.0, \{duration\}]. \newline
\textbullet~n\_occurrences\_hint: ``multi'' (2+ intervals) or ``single'' (1). \newline
\textbullet~verified: true on every emitted QA. \newline
\emph{Verification, balance, abstain:} seek to each interval before writing; aim for roughly half multi and half single; return [] if the audio is silent while captions describe clear events, if captions describe events not heard at all, or if captions and audio disagree on more than roughly half the events. \\ \addlinespace
\textbf{Output format.} Return ONLY a JSON array of \{K\} objects (or [] to abstain). No prose, no markdown fences. Example: \texttt{[\{"query": "A steady, coarse scraping of sandpaper on wood dominates the foreground", "intervals": [[0.0, 12.5],[18.0, 30.0]], "n\_occurrences\_hint": "multi", "verified": true\}, \{"query": "A brief, muffled cough is barely audible in the background", "intervals": [[14.2, 14.9]], "n\_occurrences\_hint": "single", "verified": true\}]}. \\ \addlinespace
\textbf{User-turn template} (audio attached; appended below the instructions above): \newline
\texttt{AUDIO CLIP: (attached --- please listen)} \newline
\texttt{duration\_sec: \{duration\}} \par\vspace{2pt}
\emph{COARSE CAPTIONS (from an automated model --- verify against the audio):} \newline
\texttt{\{events\_block\}} --- one line per caption: \texttt{- [\textless on\textgreater{} - \textless off\textgreater] \textless caption text\textgreater} \par\vspace{2pt}
\texttt{Now produce the JSON array with exactly \{K\} queries, or [] if you must abstain.} \\
\end{longtable}
\endgroup